\documentclass[printer]{tJOT2e}

\usepackage{psfig}
\usepackage{epsfig}

\newcommand{\bfi}{\bfseries\itshape}
\newcommand{\be}{\begin{equation}}
\newcommand{\en}{\end{equation}}
\newcommand{\bea}{\begin{eqnarray}}
\newcommand{\ena}{\end{eqnarray}}
\newcommand{\ovl}{\overline}
\newcommand{\half}{\frac{1}{2}}
\newcommand{\pr}{\partial}
\newcommand{\ub}{{\overline{u}}}
\newcommand{\uub}{{\overline{{\bf{u}}}}}
\newcommand{\uu}{{\bf{u}}}
\newcommand{\pb}{{\overline{p}}}

\begin{document}

\doi{10.1080/14685240xxxxxxxxxxxxx}
\issn{1468-5248}
\issnp{} \jvol{00} \jnum{00} \jyear{2005} \jmonth{April}

\markboth{Leray and LANS$-\alpha$ modelling of turbulent mixing}{Leray and LANS$-\alpha$ modelling of turbulent mixing}

\title{Leray and LANS$-\alpha$ modelling of turbulent mixing}

\author{Bernard J. Geurts\footnote[7]{To whom correspondence should be addressed (b.j.geurts@utwente.nl)} \\~\\
Multiscale Modelling and Simulation, Faculty EEMCS, University of Twente,\\ P.O. Box 217, 7500 AE Enschede, The Netherlands,~~and \\ Fluid Dynamics Laboratory, Department
of Applied Physics, Eindhoven University of Technology, P.O. Box 513, 5600 MB Eindhoven, The Netherlands \\~\\
Darryl D. Holm \\~\\
Computational and Computer Science Division, MS D413, Los Alamos National Laboratory, Los Alamos, NM 87545, USA,~~and \\ Mathematics Department, Imperial College London,
SW7 2AZ, London UK}

\maketitle

\begin{abstract}
Mathematical regularisation of the nonlinear terms in the Navier-Stokes equations provides a systematic approach to deriving subgrid closures for numerical simulations of turbulent flow. By construction, these subgrid closures imply existence and uniqueness of strong solutions to the corresponding modelled system of equations. We will
consider the large eddy interpretation of two such mathematical regularisation principles, i.e., Leray and LANS$-\alpha$ regularisation. The Leray principle introduces a {\bfi smoothed transport velocity} as part of the regularised convective nonlinearity. The LANS$-\alpha$ principle extends the Leray formulation in a natural way in which a {\bfi filtered Kelvin circulation theorem}, incorporating the smoothed transport velocity, is explicitly satisfied. These regularisation principles give rise to implied subgrid closures which will be applied in large eddy simulation of turbulent mixing. Comparison with filtered direct numerical simulation data, and with predictions obtained from popular dynamic eddy-viscosity modelling, shows that these mathematical regularisation models are considerably more accurate, at a lower computational cost. Particularly, the capturing of flow features characteristic of the smaller resolved scales improves significantly. Variations in spatial resolution and Reynolds number establish that the Leray model is more robust but also slightly less accurate than the LANS$-\alpha$ model. The LANS$-\alpha$ model retains more of the small-scale variability in the resolved solution. This requires a corresponding increase in the required spatial resolution. When using second order finite volume discretisation, the potential accuracy of the implied LANS$-\alpha$ model is found to be realized by using a grid spacing that is not larger than the length scale $\alpha$ that appears in the definition of this model.
\\~\\

\noindent
This paper is associated with the focus-issue {\it{Cascade Dynamics: Fundamentals and Modelling.}}

\end{abstract}

\section{Introduction}

\subsection{The Large Eddy Simulation (LES) approach for modelling turbulent flow}

The need for accurate and efficient prediction of turbulent flow has increased in priority during the past few decades, stimulated by the scientific as well as the practical significance of this field of interest. Various computational modelling strategies to address this need have been put forward. Conceptually, the simplest approach is to discretise the governing Navier-Stokes equations and resolve all dynamically relevant length scales that are contained in an unsteady turbulent solution. This direct numerical simulation (DNS) approach has proven to be an essential point of departure for understanding fundamental properties of turbulence and to provide a reliable under-pinning of theoretical and modelling approaches. However, the requirement of numerically resolving flow-features down to the Kolmogorov length-scale poses severe restrictions on direct numerical simulation. In fact, current computational resources restrict applications of DNS to turbulent flow of modest complexity, i.e.,  to low Reynolds numbers. As a consequence, much research has been directed to develop simulation strategies that are computationally much less demanding and at the same time provide sufficient accuracy for specific applications. This necessarily implies a coarsening of the flow description which is usually formalised in terms of an averaging process that explicitly reduces the dynamic importance of the smaller features in a flow, by modelling their effects in terms of the resolved features. Different reduced descriptions that have been proposed in the literature may be distinguished by their specific underlying averaging process. The well-known Reynolds averaged Navier-Stokes (RaNS) formulation is usually defined with reference to either a long-time averaging or an ensemble averaging. More recent is the so-called large eddy simulation (LES) which is based on the application of a low-pass spatial averaging to the Navier-Stokes equations.

The application of an averaging process to the Navier-Stokes equations imposes an external control over the dynamic importance of the small-scale flow-features. Usually, the averaging is supposed to commute with temporal and spatial derivatives. This preserves the momentum-conservation form of the governing equations, while introducing the so-called `closure' terms. These closure terms are associated with averaging the convective nonlinearity of the Eulerian equations and they give rise to momentum fluxes that cannot be fully evaluated in terms of the averaged solution alone. In RaNS context, this closure problem is defined in terms of the Reynolds stress tensor while in LES the turbulent stress tensor arises. In this paper we will restrict ourselves
to LES. In order to arrive at a meaningful formulation for the prediction of the averaged solution, the closure problem must be resolved by introducing a model for the turbulent stress tensor which approximates the dynamic consequences of the small-scale flow-features in terms of operations on the averaged solution alone. So, while the main virtue of the LES approach is to provide external control over the dynamic complexity of the simulation model, its main challenge is to provide an accurate closure for the turbulent stresses.

A significant portion of all LES literature is devoted to developing accurate models for the closure problem that are introduced by spatial filtering. Popular low-pass filters
used in LES, such as the top-hat or Gaussian filters, may be characterised by a length-scale parameter that represents the `width' $\Delta$ of the filter. Application
of such filters suppresses the contributions of flow-features that are more localised than $\Delta$ while the larger, more energy-containing scales are virtually unaffected
by the filtering. This implies that the closure problem for the turbulent stress tensor primarily involves modelling the effects of the `sub-$\Delta$' flow features on the larger flow features. Since in numerical simulations the filter-width and the spatial resolution are typically chosen to be of comparable magnitude, the closure model for the turbulent stresses is usually referred to as a {\bfi subgrid model}. 

Various suggestions have been made for obtaining acceptable subgrid models. In the absence of a comprehensive statistical theory of turbulence, empirical knowledge about small-scale turbulence is essential for the development of such subgrid models. This includes a detailed representation of the dissipative and dispersive contributions of the sub-$\Delta$ features to the dynamics of the larger features.
Moreover, guidance for proper subgrid modelling may be obtained by requiring the modelled equations to comply as much as possible with basic properties of the (filtered) Navier-Stokes equations \cite{ghosalaiaa}. In this context one may, e.g., consider: preserving Galilean transformation properties of the subgrid model \cite{speziale,horiuti},
maintaining realisability conditions in case positive filters are adopted \cite{vreman_realize}, accounting for algebraic properties of the turbulent stress tensor as basis for dynamic subgrid modelling \cite{germano}, incorporating resolved small-scale information by allowing (approximate) inversion of the LES filter \cite{geurts_inverse}, or combining filter inversion with the dynamic procedure \cite{kgvg}, just to name a few of these `mathematical' modelling options.

An extensive account of modelling practices in LES may be found in \cite{sagautbook}. The variations among the results arising from the different subgrid models that have been proposed appears inconsistent with the basic premise of LES that small-scale turbulence may be assumed to be `universal'. Such `universality' would suggest that the characteristic features of the large scales in a flow problem should show a degree of insensitivity to the properties of its small-scale flow features, provided these features are modelled in the proper universality class. Correspondingly, accurate subgrid models should exist which are simple, in the sense that one should be able to formulate such models in terms of basic properties of the turbulent flow without involving, for example, details related to the geometry of the flow domain. If universality is a proper guiding principle for subgrid modelling, then to the extent the results of the numerical simulations depend on these modelling steps, the intuitive and empirical modelling steps that have primarily been considered in the LES literature may not belong to the proper universality class. Regardless of whether the premise of universality holds, we suggest a alternative systematic approach based on returning to the turbulence closure problem, concentrating on the mathematical and physical aspects of the Navier-Stokes equations, and seeking the simplest subgrid model which is still consistent with these fundamental aspects.

Turbulence simulations based directly on the governing Navier-Stokes equations are not only hampered by the complexity of turbulent solutions. In addition, there exist   outstanding mathematical issues concerned with existence, uniqueness and regularity of solutions to these flow-equations. These issues address the fundamental computability of turbulent flow. Thus, they introduce uncertainty into the fields of (dedicated) numerical methods and subgrid modelling of turbulence. In contrast, one may limit possible closure strategies such that the final formulation at least is guaranteed to possess a unique solution with {\it a priori} known smoothness properties. Moreover, it appears natural to require that the solutions of the unfiltered Navier-Stokes equations would be recovered in the mathematical limit in which the filter-width is sequentially reduced to zero. Recently, analysis established such properties for the well-known Smagorinsky model \cite{smagorinsky,berselli}. However, this subgrid model represents the dynamics of the small turbulent scales in terms of a nonlinear eddy-viscosity only, and this limitation does not do justice to all intricacies of the spatially filtered convective fluxes. Moreover, from experience gathered in LES using this subgrid model, it has become clear that the predictions of the filtered Navier-Stokes solution by the Smagorinsky model are in many cases not accurate enough to be of practical relevance \cite{vremanjfm}.

\subsection{Averaging, filtering and closure: LES and Lagrangian averaging (LA)}
We begin motivating our selection of a class of reduced flow descriptions for which  existence and regularity properties are available, by recalling a historical result for the Navier-Stokes equations which regularised their convective
nonlinearity. In mathematical analysis, this approach was
pioneered in the classic work by Leray \cite{leray} who
introduced a {\bfi smoothed transport velocity} in the convective
nonlinearity. In detail, Leray replaced ${\bf{u}}\cdot \nabla
{\bf{u}}$ in the Navier-Stokes equations by 
${\overline{\bf{u}}}\cdot \nabla {\bf{u}}$ where
the over-line on ${\overline{\bf{u}}}$ denotes filtering of the velocity
${\bf{u}}$. Both the velocities ${\overline{\bf{u}}}$ and ${\bf{u}}$
should be regarded as regularised quantities. 
For a low-pass filter whose kernel decreases
properly to zero for large values of its argument, Leray
established uniqueness and regularity of the solution ${\bf{u}}$ of the
Leray-regularised equations, as well as convergence to a
Navier-Stokes solution as the filter width tended to
zero. This regularisation principle maintains the conservative 
structure of the filtered equations and represents a
mathematically-based approach for obtaining reduced
descriptions of turbulent flow. As such, 
Leray's regularisation principle may bridge the gap
between practical requirements posed by LES and the high
level of mathematical rigour required to guarantee systematic
progress in turbulence analysis. The main practical question,
of course, is whether or not this coarsening of the flow
description leads to accurate predictions of the smoothed
flow. One of the goals of the present paper is to resolve this issue  
favourably in the context of turbulent mixing in a shear layer.

In comparison to Leray's regularisation principle, the
framework of the Lagrangian averaged Navier-Stokes$-\alpha$
(LANS$-\alpha$) equations provides an alternative systematic
method for modelling the mean circulatory effects of
small-scale turbulence, while maintaining the mathematical
properties that guarantee existence of a unique, regular
solution and a finite dimensional global attractor \cite{fht2001,ChHoOlTi2004}. The
inviscid Lagrangian averaged Euler$-\alpha$ equations were
originally derived as Euler-Poincar\'e equations in the
framework of Hamilton's principle for geometric fluid
mechanics \cite{HoMaRa1998}. The corresponding LANS$-\alpha$
model for turbulent flow may also be obtained more directly, by
applying Lagrangian filtering to the Kelvin circulation
theorem for the incompressible motion of a fluid with
viscosity. As a consequence, the equations for the
LANS$-\alpha$ model are obtained, and these generalise the
Leray regularisation. The LANS$-\alpha$ solutions also
converge to the unfiltered Navier-Stokes solutions as the
filter width tends to zero. However, because the filtered
Kelvin theorem is built into this turbulence modelling
approach, it is also suitable for modelling turbulent flows
in a rotating frame of reference and with buoyancy, as required in
weather and climate modelling. The present paper compares the performance of the
LANS$-\alpha$ model with the Leray model in numerical simulations of turbulent
mixing and investigates the improvements in its predictions 
relative to the required computational effort. These findings
are compared to traditional subgrid models in simulations obtained using the popular dynamic eddy-viscosity model \cite{germano_dynmod}.

\subsection{Outline of the paper}
This paper is organised as follows. Section \ref{regular} presents and analyses regularisation modelling for large eddy simulation of turbulent flow. The
Leray and LANS$-\alpha$ models will be derived and discussed. Section \ref{picsmix} describes the turbulent mixing layer and the numerical methods adopted. The main features of the developing transitional and turbulent flow as captured by the regularisation and dynamic subgrid models are presented to characterise the global trends. Section \ref{leraymix} focuses on the quantitative predictions of turbulent mixing obtained with the Leray model. Mean flow, fluctuating quantities and spectral properties of the flow will be considered in order to assess the accuracy of the Leray model in some detail, both at low and at high Reynolds numbers. Section \ref{nsalpha} is dedicated to improvements in the predictions that arise from the $\alpha$-extensions of the Leray principle. It is shown that small-scale variability is restored to some degree in the LANS$-\alpha$ model, which goes at the expense of more strict requirements on the spatial resolution. Concluding remarks are collected in section \ref{concl}.

\section{Regularisation modelling for large eddy simulation}
\label{regular}

In this section we relate mathematical regularisation of the convective fluxes to implied subgrid models for LES. First, in subsection \ref{filterles} we review the filtering
approach to large eddy simulation. Moreover, we describe the common phenomenological treatment of the closure problem in which dissipative - or dispersive features of the
turbulent stresses are represented by (dynamic) eddy-viscosity or similarity modelling respectively. This provides the context for discussing the basic mathematical regularisation
strategy. The Leray and LANS$-\alpha$ regularisation principles provide central examples of this approach and the corresponding `implied' subgrid models are discussed in
subsection \ref{regclose}. Finally, a simple Fourier-mode analysis of the regularisation models is collected in subsection \ref{fourier} and compared with results obtained for
standard similarity models.

\subsection{Filtering approach to large eddy simulation}
\label{filterles}

Filtering the Navier-Stokes equations requires a low-pass spatial filter $L$. Often, a convolution filter is adopted which in one spatial dimension associates the filtered
velocity $\ub$ with the unfiltered velocity $u$ through
\begin{equation}
\ub=L(u)=\int_{-\infty}^{\infty} G(x-\xi,\ell) u(\xi) ~d\xi
\end{equation}
with normalised filter-kernel $G(z,\ell)$. This filter-kernel is characterised by an externally specified length-scale parameter $\ell$ which defines the effective filter-width
$\Delta$, e.g., as \cite{bosgeurts_api1}
\begin{equation}
\frac{1}{\Delta}= \int_{-\infty}^{\infty} G^{2}(z,\ell) dz =\frac{1}{2 \pi} \int_{-\infty}^{\infty} |H(k,\ell)|^{2}~ dk
\label{fildef}
\end{equation}
where $H(k,\ell)$ is the Fourier-transform of the filter-kernel. This definition applies to all kernels that are square integrable such as the popular top-hat filter, the
spectral cut-off filter or the Gaussian filter \cite{geurtsbook2003}. Other definitions proposed in literature (see \cite{sagautbook} for an overview) are more restricted in
their applicability to different filters.

For incompressible fluids, the application of the filter $L$ to the continuity and Navier-Stokes equations leads to:
\be
\pr_{j}\ub_{j}=0\,,\quad\hbox{and}\quad
\pr_{t}\ub_{i}+\pr_{j}(\ub_{j}\ub_{i})+\pr_{i}\pb-\frac{1}{Re}\pr_{jj}
\ub_{i}= {{-\pr_{j}({\overline{u_{i}u_{j}}}-\ub_{j}\ub_{i})}}
\label{filnseqs}
\en
Here $\pr_{t}$ (resp. $\pr_{j}$) denotes partial differentiation with respect to time $t$ (resp. spatial coordinate $x_{j}$). Summation over repeated indices is implied. The
component of the filtered velocity in the $x_{j}$ direction is $\ub_{j}$ and $\pb$ is the filtered pressure. Finally, $Re$ denotes the Reynolds number of the flow.

In formulation (\ref{filnseqs}) of the {\bfi LES-template} \cite{geurtsbook2003} we recognise the application of the {\bfi Navier-Stokes operator} to the filtered solution
$\{\ub_{j},\pb\}$ on the left-hand side. On the right-hand side, the terms expressing the central closure problem are collected in the divergence of the turbulent stress tensor
$\tau_{ij}={\overline{u_{i}u_{j}}}-\ub_{i}\ub_{j}$. This tensor cannot be calculated from the filtered solution alone. Hence, one of the aims in the development of large eddy
simulation is the effective capturing of the primary dynamical effects of $\tau_{ij}$ in terms of model tensors that may be evaluated in terms of operations on the filtered
solution alone.

In the absence of a comprehensive theory of how the small scales of turbulence influence its large scales, empirical knowledge about modelling $\tau_{ij}$ is essential, but it is still rather incomplete. Usually, LES subgrid models are proposed either on the basis of their presumed dissipative nature, or in view of the scale-similarity property of $\tau_{ij}$ in an inertial range (\cite{meneveaukatz}). As further guidance in the construction of suitable models, one may attempt to incorporate information associated with mathematical properties of the modelling problem such as realisability conditions (\cite{vreman_realize}), algebraic identities (\cite{germano}) or approximate inversion of the filter (\cite{geurts_inverse}, \cite{stolzadams}). While realisability conditions may impose bounds on certain model parameters, the incorporation of algebraic properties such as Germano's identity, possibly combined with filter-inversion \cite{kgvg}, has led to a successful class of {\bfi dynamic subgrid models}. These dynamic models represent the current state of the art, at least for flows away from solid boundaries.

Much of turbulence phenomenology is captured in the Kolmogorov picture, in which kinetic energy cascades in an average sense through an {\bfi inertial range} toward ever smaller scales, until its flow features are sufficiently localised that they may be effectively dissipated by viscosity \cite{kolmogorov,frisch}. The dynamics is dominated by viscosity for flow features whose sizes (length scales) are smaller than the Kolmogorov length $\eta_{K}$. When a filter of filter-width $\Delta \gg \eta_{K}$ is applied to a turbulent solution, (virtually) all molecular dissipation associated with the small scales is removed. In LES, the filter width is commonly chosen to be within a presumed inertial range in which the dynamics is dominated by convection. The continuous cascading of energy through the inertial range is then usually balanced by an ``extra'' eddy-viscosity contribution. This may yield a computational model that at least retains accurate large-scale information in the filtered solution. Dissipation of turbulent kinetic energy was first parameterised by using the {\bfi Smagorinsky model} \cite{smagorinsky} in which one approximates:
\begin{equation}
\tau_{ij} \rightarrow m_{ij}^{S}=-(C_{S}\Delta)^{2} |S| S_{ij}
\end{equation}
where $C_{S}$ denotes Smagorinsky's constant, $S_{ij}=\pr_{i}\ub_{j}+\pr_{j}\ub_{i}$ is the rate of strain tensor and $|S|^{2}=S_{ij}S_{ij}$ is its magnitude.

Besides the modelling of dissipation, a large volume of literature on the phenomenological treatment of the turbulent stress tensor is based on the assumed {\bfi scale-similarity properties} of this tensor when the filter-width is taken to be somewhere within the inertial range \cite{meneveaukatz}. More precisely, this suggests that approximate models for $\tau_{ij}$ can be obtained by applying its definition to an appropriate operation on the filtered solution. We may express $\tau_{ij}$ as the commutator of the spatial filter and the product
operator \cite{geurts_inverse,kgvg}, i.e.,
\begin{equation}
\tau_{ij}={\overline{u_{i}u_{j}}}-\ub_{i}\ub_{j}=L(\Pi_{ij}({\bf{u}}))-\Pi_{ij}(L({\bf{u}}))=[L,\Pi_{ij}]({\bf{u}})
\end{equation}
where the product operator $\Pi_{ij}({\bf{u}})=u_{i}u_{j}$. The similarity aspects of the closure problem in the inertial range were first parameterised by the {\bfi Bardina model}
\cite{bardina}. In this model one puts
\begin{equation}
\tau_{ij} \rightarrow m_{ij}^{B}={\overline{\ub_{i}\ub_{j}}} -{\overline{\ub}}_{i}{\overline{\ub}}_{j} = [L,\Pi_{ij}]({\overline{\bf{u}}})
\end{equation}
Thus, the Bardina model applies the definition of $\tau_{ij}$ to the available filtered velocity.

Extensions have been proposed in which the filtered velocity ${\overline{\bf{u}}}$ is replaced by an approximate reconstruction of the unfiltered solution. In fact, for so-called graded filters $L$ (e.g., including the top-hat and Gaussian filters, but excluding the spectral cut-off filter), operating on ${\overline{\bf{u}}}$ with a formal approximation
of the inverse $L^{-1}$, leads to a partial reconstruction of the unfiltered solution, at least where (most of) the resolved scales are concerned. This reconstructed solution may
be used to define generalised similarity models \cite{geurts_inverse}. This approach  forms the basis for the approximate de-convolution models developed in \cite{stolzadams}. These (generalised) similarity models characterise the dispersive effects of the small-scale turbulence on the resolved flow features. When smooth test-solutions are considered and high-order approximate inversion, it may be shown that the difference between $\tau_{ij}$ and the generalised model tensor scales as $\Delta^{m}$ \cite{geurts_inverse,geurtsbook2003} where the power $m$ is controlled by the quality of the inversion. Of course, this scaling only holds in the mathematical limit $\Delta
\rightarrow 0$; actual simulations with these models indicate the need for additional smoothing, e.g., through extra eddy-viscosity contributions. This is usually achieved in the context of the dynamic procedure to which we turn next. 

\subsection{Dynamic mixed models}
The Smagorinsky and similarity models separately describe important intuitive features of the turbulent stresses. However, these models are well known to be seriously flawed in
their own ways. The Smagorinsky model displays low levels of correlation with $\tau_{ij}$ and often leads to excessive dissipation, especially near solid walls and in laminar
flows with large gradients. This may even hinder a modelled flow from going through a complete transition to turbulence \cite{vremanjfm}. The similarity model of Bardina is
known to display high correlation \cite{meneveaukatz}, but it fails to provide effective dissipation of energy and it may give rise to unrealistically high levels of small scale
fluctuations in the solution. For these reasons, so-called {\bfi mixed models} have been proposed which combine similarity with eddy-viscosity models
\cite{vremangeurtskuertenformulation}. As an example, a basic mixed model is $\tau_{ij} \rightarrow m_{ij}^{M}=m_{ij}^{B}-C_{d}\Delta^{2}|S|S_{ij}$ in which $C_{d}$ denotes the
eddy coefficient. The central problem that now arises is how this eddy coefficient should be specified in accordance with the evolving flow. A well-known and elegant way to
approach this without unduly introducing ad hoc parameters is based on Germano's identity \cite{germano_dynmod} which provides the basis for the dynamic subgrid modelling
procedure.

In {\bfi dynamic models}, the eddy-viscosity is intended to reflect local instantaneous turbulence levels. One starts from {\bfi Germano's identity}: 
\be
T_{ij}-\widehat{\tau_{ij}}=R_{ij}
\,,
\en
where
\be
T_{ij} = \widehat{\ovl{ u_i u_j}}-\widehat{\ovl{u}}_i \widehat{\ovl{ u}}_j
\quad\hbox{ and }\quad
R_{ij} = {\widehat{(\ovl{ u}_i \ovl{u}_j)}} -\widehat{\ovl{u}}_i \widehat{\ovl{u}}_j
\,.
\en
Here, in
addition to the basic LES-filter $\ovl{(\cdot)}$ of width ${{\Delta}}$, a so-called explicit {\bfi test filter} $\widehat{(\cdot)}$ of width ${\widehat{\Delta}}$ is introduced. The
only external parameter to be specified is the ratio of filter widths, which is often assigned as $\kappa={\widehat{\Delta}}/{{\Delta}}=2$. The implementation of the dynamic
procedure starts by assuming a (mixed) model $m_{ij}= a_{ij}(\uub)+c b_{ij}(\uub)$ for $\tau_{ij}$ and $M_{ij}=A_{ij}+C B_{ij}$ for $T_{ij}$ where
$A_{ij}=a_{ij}({\widehat{\uub}})$, $B_{ij}=b_{ij}({\widehat{\uub}})$. Here $a_{ij}$ and $b_{ij}$ express assumed basic models, e.g., similarity or eddy-viscosity models, and the
assumption $C \approx c$ is usually invoked. If we introduce ${\cal A}_{ij}=A_{ij}-\widehat{a_{ij}}$, ${\cal B}_{ij}= B_{ij} - \widehat{b_{ij}}$ and use the common approximation
${\widehat{(cb_{ij})}} \approx c ({\widehat{b_{ij}}})$, insertion in Germano's identity yields ${\cal A}_{ij}+c{\cal B}_{ij}=R_{ij}$. This relation should hold for all
tensor-components, which of course is not possible for a single scalar coefficient field $c$. To resolve this situation we introduce a linear averaging operator
$\langle \cdot \rangle $ and define the {\bfi Germano residual} by
\be
\varepsilon(c)= \langle \frac12 \{(R_{ij}-{\cal A}_{ij})-c{\cal B}_{ij}\}^{2} \rangle
\label{germanoresidual}
\en
The averaging can involve integration over parts of the flow domain \cite{vremanjfm}, over past history of the solution or include some Lagrangian averaging
\cite{meneveaulagrangian}. We hence obtain an optimality condition for $c$ by requiring the variation of $\varepsilon$ to vanish. We can solve the local coefficient as
\cite{lilly}
\be
c=\frac{\langle (R_{ij}-{\cal A}_{ij}){\cal B}_{ij}\rangle}{\langle {\cal B}_{ij}{\cal B}_{ij}\rangle}
\label{dyncoef}
\en
where we assumed $\langle c fg \rangle \approx c \langle f g \rangle$. In order to prevent numerical instability caused by negative values of the eddy-viscosity, the dynamic
coefficient is also artificially set to zero wherever (\ref{dyncoef}) returns negative values. This is referred to as `clipping'. For further details we refer to
\cite{vremanthesis,geurtsbook2003}.

Dynamic subgrid models have contributed in many ways to the understanding of turbulent flow in complex situations. However, dynamic models are also known to be hampered by a
number of drawbacks. First, the dynamic procedure is quite expensive in view of the relatively large number of additional explicit filter operations that need to be included.
Moreover, the implementation contains various {\it{ad hoc}} elements or inaccurate assumptions such as the independence of the dynamic coefficient on the filter-level, or the well-known `clipping' of negative eddy-viscosity, required to ensure stability of a simulation. The achieved accuracy in actual simulations remains quite limited, e.g., due to shortcomings in the assumed base models. Since the dynamic approach does not contain external parameters other than the ratio between the width of the explicit test-filter and the basic LES filter, there is no chance of improving the predictions by `tinkering' with parameters. Finally, an extension to flows involving complex physics and/or developing in complex flow-domains is difficult since no systematic framework exists for this purpose. For these reasons, an alternative modelling approach is summoned and we turn to the recently proposed regularisation modelling (\cite{gh2003}) in the next subsection.

\subsection{Regularisation strategy to subgrid closure}
\label{regclose}

In this subsection we consider two regularisation principles for the Navier-Stokes equations and derive the associated subgrid models in case the basic filter $L$ has a formal
inverse $L^{-1}$ \cite{gh2003}. We consider {\bfi Leray regularisation} \cite{leray} and the {\bfi Lagrangian averaged Navier-Stokes-$\alpha$ (LANS$-\alpha$)} approach \cite{fht2001}.

The mathematical regularisation of the Navier-Stokes equations which we pursue here involves a direct and explicit alteration of the nonlinear convective terms. In the context of this paper, this provides a systematic framework for {\it{deriving}} a subgrid model which is in sharp contrast with traditional phenomenological subgrid modelling. As sketched in the previous subsection, such phenomenological subgrid modelling achieves the desired smoothing of a turbulent flow only indirectly by quite `independently' introducing a subgrid model for the turbulent stresses.

Phenomenological subgrid models are usually only loosely connected to the basic LES filter $L$ that was used to arrive at the unclosed equations. For example, eddy-viscosity models at best explicitly incorporate the width of the filter, and the modelled system does not contain any further information that characterises the shape of the particular filter. For similarity models there is a more direct relation with the explicit filter, although properties of the basic LES filter are never actually introduced, other than that
the filter needs to be of convolution type. This `independence' of the specific filter also holds for the dynamic procedure which only requires explicit specification of the
test-filter. So, although the filter $L$ is the only element that defines the relation between DNS and LES, as well as the detailed properties of the subgrid stresses, the
actual phenomenological modelling does not reflect this central role of $L$. Moreover, introducing dissipation by eddy-viscosity does not do justice to the intricacies of
turbulent transport phenomena; it can at best perhaps characterise effective statistical properties of the kinetic energy dynamics. These fundamental considerations are fully
respected in the regularisation principles that are considered next.

\paragraph{Approach} A mathematical modelling approach for large eddy simulation can be obtained by combining a {\bfi regularisation principle} with an explicit filter and its formal inversion
\cite{gh2003}. Historically, the first example of a smoothed flow description in this category is the Leray regularisation \cite{leray}. Although this regularisation was
introduced for entirely different reasons, we may reinterpret the Leray proposal in terms of its implied subgrid-model. This provides a direct connection with large eddy
simulation.

The application of a specific filter in combination with a mathematical regularisation of Navier-Stokes equations allows one to incorporate properties from first
principles into the modelled equations. Appropriate mathematical regularisation results in a system of equations that is guaranteed to have a unique solution with suitable
smoothness. The dynamics are characterised by an attractor of finite dimensions which has very favourable consequences for the computability of the solution, compared to the
unfiltered Navier-Stokes system. Moreover, the momentum-conservation structure of the equations is retained.

\subsubsection{Leray regularisation}
In Leray regularisation, one alters the convective fluxes into $\ub_{j}\pr_{j}u_{i}$, i.e., the solution $\uu$ is convected with a smoothed velocity $\uub$. Consequently, the
nonlinear effects are reduced by an amount governed by the smoothing properties of the filter operation, $L$. The governing Leray equations are \cite{leray,ChHoOlTi2004}
\begin{equation}
\pr_{j}\ub_{j}=0~~;~~\pr_{t}u_{i}+\ub_{j}\pr_{j}u_{i}+\pr_{i}p
-\frac{1}{Re}\Delta u_{i}=0
\label{lerayeqs}
\end{equation}
Leray solutions possess global existence and uniqueness with proper smoothness and boundedness, whose demonstration depends on the balance equation for $\int\!|\mathbf{u}|^2\,d\,^3x$.
Based on the Leray equations (\ref{lerayeqs}) we may eliminate $\uu$ by assuming $\uub=L(\uu)$ and $\uu=L^{-1}(\uub)$. For convolution filters one has, e.g., $\pr_{t} u_{i}=
\pr_{t}(L^{-1}(\ub_{i}))=L^{-1}(\pr_{t}\ub_{i})$ and the nonlinear terms can be written as ${{{\overline{u}}_{j}}}\pr_{j}(u_{i}) = \pr_{j}({{{\overline{u}}_{j}}}u_{i}) =
\pr_{j}({{{\overline{u}}_{j}}}L^{-1}(\ub_{i}))$. Consequently, one may readily obtain:
\begin{equation}
L ^{-1} \Big(\pr_{t}\ub_{i} + \pr_{j}(\ub_{j}\ub_{i}) + \pr_{i} \pb - \frac{1}{Re}\Delta \ub_{i} \Big) =
{{- \pr_{j} \Big( {{{\overline{u}}_{j}}}L^{-1}(\ub_{i}) - L^{-1}(\ub_{j}\ub_{i}) \Big)}}
\end{equation}
This may be recast in terms of the LES template as:
\begin{equation}
\pr_{t}\ub_{i} + \pr_{j}(\ub_{j}\ub_{i}) + \pr_{i} \pb - \frac{1}{Re} \Delta \ub_{i} ={{- \pr_{j}\Big(m_{ij}^{L}\Big)}}
\label{leray_eq}
\end{equation}
The implied asymmetric Leray model $m_{ij}^{L}$ involves both $L$ and its inverse and may be expressed as:
\begin{equation}
m^{L}_{ij}=L\Big(\ub_{j}L^{-1}(\ub_{i})\Big) -\ub_{j}\ub_{i} ={\overline{\ub_{j}u_{i}}} -\ub_{j}\ub_{i}
\label{leraymodel}
\end{equation}
The reconstructed solution $u_{i}$ is found from any formal or approximate inversion $L^{-1}$. For this purpose one may use a number of methods, e.g., polynomial inversion
\cite{geurts_inverse}, geometric series expansions \cite{stolzadams} or exact numerical inversion of Simpson top-hat filtering \cite{kgvg}.

\subsubsection{LANS$-\alpha$ regularisation by Kelvin filtering}
A regularisation principle which additionally possesses correct circulation properties may be obtained by starting from the following Kelvin theorem:
\begin{equation}
\frac{d}{dt} \oint_{\Gamma(\mathbf{u})} u_{j}~dx_{j} - \frac{1}{Re} \oint_{\Gamma(\mathbf{u})} \Delta u_{j}~ d x_{j}=0
\label{ktheorem}
\end{equation}
where $\Gamma(\mathbf{u})$ is a closed fluid loop moving with the Eulerian velocity $\mathbf{u}$. The unfiltered Navier-Stokes equations may be derived from (\ref{ktheorem}) \cite{fht2001,geurtsbook2003}. This provides some of the inspiration to arrive at an alternative regularisation principle for Navier-Stokes turbulence \cite{fht2001}. In fact, the basic
regularisation principle was originally derived by applying {\bfi Taylor's hypothesis} of frozen-in turbulence in a Lagrangian averaging framework \cite{holmphysd}. In this framework, the fluid loop is considered to move with the smoothed {\bfi transport velocity} $\uub$, although the circulation velocity is still the unsmoothed velocity, $\mathbf{u}$. That is, in (\ref{ktheorem}) we replace $\Gamma(\mathbf{u})$ by $\Gamma({\overline{\mathbf{u}}})$; so the material loop $\Gamma$ moves with the filtered transport velocity. From this filtered Kelvin principle, we may obtain the {\bfi Euler-Poincar\'e equations} governing the smoothed solenoidal fluid dynamics, with $\partial_j\ub_{j}=0$ and \cite{HoMaRa1998}
\begin{equation}
\pr_{t}u_{j}+\ub_{k}\pr_{k}u_{j}+u_{k}\pr_{j}\ub_{k}+\pr_{j}p - \pr_{j}(\frac12 \ub_{k}u_{k}) -\frac{1}{Re}\Delta u_{j}=0
\label{epeq}
\end{equation}
Comparison with the
Leray regularisation principle in (\ref{lerayeqs}) reveals two additional terms in (\ref{epeq}). These terms guarantee the regularised flow to be consistent with the modified
Kelvin circulation theorem in which $\Gamma(\mathbf{u})\to\Gamma({\overline{\mathbf{u}}})$. For LANS$-\alpha$ the analytical properties of the regularised solution are based on the
energy balance for $\int\!\mathbf{u}\cdot\,L(\mathbf{u})\,d\,^3x$.

The Euler-Poincar\'{e} equations (\ref{epeq}) can also be rewritten in the form of the LES template. The extra terms that arise in (\ref{epeq}) give rise to additional terms in
the implied subgrid model:
\begin{equation}
\pr_{t}\ub_{i} + \pr_{j}(\ub_{j}\ub_{i}) + \pr_{i} \pb - \frac{1}{Re} \Delta \ub_{i}
= -\pr_{j}\Big({\overline{\ub_{j}u_{i}}}-\ub_{j}\ub_{i}\Big) -\frac12 \Big( {\overline{u_{j}\pr_{i}\ub_{j}-\ub_{j}\pr_{i}u_{j}}}\Big)
\label{genalpha}
\end{equation}
We observe that the Leray model (\ref{leraymodel}) reappears as part of the implied LANS$-\alpha$ subgrid model on the right-hand side of (\ref{genalpha}). Compared to the Leray
model, the additional second term in the LANS$-\alpha$ model takes care of recovering the Kelvin circulation theorem for the smoothed solution. This formulation is given in terms
of a general filter $L$ and its inverse. After some further rewriting it may be shown that this model can be formulated in conservative form, i.e., a tensor $m_{ij}^{\alpha}$ can
be found such that the right hand side of (\ref{genalpha}) can be written as $-\pr_{j}m_{ij}^{\alpha}$. We illustrate this next for a particular filter.

The subgrid model presented in (\ref{genalpha}) can be specified further in case the filter $L$ has the Helmholtz operator as its inverse, i.e., $ {u}_i= L^{-1}(\ub_{i})=
(1-\alpha^{2} \pr_{jj}) \ub_{i} = {\rm He}_{\alpha}(\ub_{i})$. Then we recover the original LANS$-\alpha$ equations \cite{fht2001}. The LANS$-\alpha$ model derives its name from the
length-scale parameter {\mbox{$\alpha\approx \Delta/5$}} \cite{geurtsholm2002}. After some rewriting, the following parameterisation for the turbulent stress tensor is obtained
\cite{domaholm}:
\be
m^{\alpha}_{ij}=\alpha^{2}{\rm He}_{\alpha}^{-1}\Big(\pr_{k}\ub_{i}~\pr_{k}\ub_{j}+\pr_{k}\ub_{i}~ \pr_{j}\ub_{k} -\pr_{i}\ub_{k}~\pr_{j}\ub_{k} \Big)
\label{alpha3-model}
\en
The first term on the right-hand side is the {\it Helmholtz-filtered} tensor-diffusivity model \cite{geurtsbook2003}. The second term combined with the first term, corresponds
to Leray regularisation using Helmholtz inversion as filter. The third term completes the {\mbox{LANS$-\alpha$}} model and maintains Kelvin's circulation theorem. In
(\ref{alpha3-model}) an inversion of the Helmholtz operator ${\rm He}_{\alpha}$ is required which implies application of the {\bfi exponential filter} \cite{germanodifffilters}.
However, since the Taylor expansion of the exponential filter is identical at quadratic order to that of the top-hat or the Gaussian filters, one may approximate
${\rm He}_{\alpha}^{-1}$, e.g., by an application of the explicit top-hat filter, for reasons of computational efficiency \cite{geurtsholm2002}.

\paragraph{Relation to LES} Although the regularised turbulence equations (\ref{leray_eq}) and (\ref{genalpha}) are formally similar to LES equations, they arose from different principles, as sketched
above. Through the combination of an explicit filter and its inversion, the regularisation principle allows a systematic derivation of the implied subgrid-model. This resolves
the closure problem consistent with the adopted filter. Even though the Leray and LANS$-\alpha$ formulations may, in retrospect, be interpreted in terms of implied subgrid-models,
the regularisation equations were not obtained by applying the LES filtering method. Instead, this approach to modelling turbulence from the viewpoint of mathematical
regularisation is an {\it alternative} to LES.

In the next subsection we will adopt simple Fourier-mode analysis in one spatial dimension, in order to illustrate some basic properties of the regularisation models, before assessing these models in actual large eddy simulations of turbulent mixing in the next sections.

\subsection{Fourier-analysis of regularisation models}
\label{fourier}

In order to illustrate the dynamic effects of some of the subgrid models introduced in either of the previous subsections, we will first investigate the subgrid flux due to a
single Fourier-mode. Specifically, we select $u=\sin(kx)$ which implies $\ub=H(k\Delta) u$ for convolution filters whose kernel $G$ has Fourier-transform $H$. Substitution
in the definition of the turbulent stress yields
\begin{equation}
\tau={\overline{u^{2}}}-{\overline{u}}^{2}= \frac{1}{2}\Big(1-H^2(k\Delta)\Big)+\frac{1}{2}\Big(H^2(k\Delta)-H(2k\Delta)\Big)\cos(2kx)
\label{fourier_tau}
\end{equation}
The corresponding flux $f_{\tau}=-\partial_{x}\tau$ may be expressed as:
\begin{equation}
f_{\tau}=k\Big(H^2(k\Delta)-H(2k\Delta) \Big) \sin(2kx)=k A_{\tau}(k\Delta) \sin(2kx)
\label{fourier_tau_flux}
\end{equation}
in which we introduced the characteristic amplitude function $A_{\tau}$ which depends on $k\Delta$ only. We return to this amplitude function momentarily and use it to
assess different subgrid models.

The various subgrid models introduced above will now be evaluated in the same manner as in (\ref{fourier_tau}) and (\ref{fourier_tau_flux}). In one spatial dimension the Leray
and LANS$-\alpha$ models yield an identical flux when applied to $u=\sin(kx)$. For the Leray model one may show after some manipulation
\begin{eqnarray}
m_{L}={\overline{u\overline{u}}} - {\overline{u}}^{2} &=& \frac{1}{2}\Big(H(k\Delta)-H^{2}(k\Delta)\Big) \nonumber \\
&+&\frac{1}{2}\Big(H^{2}(k\Delta)-H(k\Delta)H(2k\Delta)\Big)\cos(2kx)
\end{eqnarray}
with corresponding flux
\begin{equation}
f_{L}=k\Big(H^{2}(k\Delta)-H(k\Delta)H(2k\Delta)\Big)\sin(2kx)=kA_{L}(k\Delta) \sin(2kx)
\label{leray_flux}
\end{equation}
Comparing (\ref{leray_flux}) with (\ref{fourier_tau_flux}) we observe a close resemblance with the only difference due to a term $H(k\Delta)$ arising where unity appears in the
exact expression. Since the filter is assumed normalised, we have $H(0)=1$ and hence the deviations will be small provided $k\Delta \ll 1$. For larger values of $k\Delta$ the
deviations may be more significant. We return to these points shortly and first complete the expressions for the other subgrid models.

For a single Fourier mode the {\bfi Bardina similarity model} results in
\begin{eqnarray}  
m_{B}={\overline{\overline{u}^{2}}} - \overline{\overline{u}}^{2}& =& \frac{1}{2}H^{2}(k\Delta)\Big(1-H^{2}(k\Delta)\Big)\nonumber \\
&+&\frac{1}{2}H^{2}(k\Delta)\Big(H^{2}(k\Delta)-H(2k\Delta)\Big)\cos(2kx)
\end{eqnarray}
with corresponding flux
\begin{equation}
f_{B}=k H^{2}(k\Delta) \Big(H^{2}(k\Delta)-H(2k\Delta)\Big)\sin(2kx)=kA_{B}(k\Delta) \sin(2kx)
\end{equation}
We observe that in this single Fourier-mode evaluation $m_{B}=H^{2}(k\Delta) \tau$, showing close agreement as $k\Delta \ll 1$. In three spatial dimensions, the Bardina model requires a number of additional applications of the explicit filter, which adds considerably to its computational cost. An approximation to the Bardina similarity model is the tensor-diffusivity model \cite{winckelmans} which does not require the extra explicit filtering. We consider this model in more detail next.

The {\bfi tensor-diffusivity model} may be derived by truncating a formal Taylor expansion of the turbulent stress tensor \cite{vremantcfd}. We express a general one-dimensional
integral filter as:
\begin{equation}
{\overline{f}}(x)=\int_{-\infty}^{\infty} \frac{1}{\Delta}G(\frac{\xi-x}{\Delta}) f(\xi) d\xi =\int_{-\infty}^{\infty} G(\eta) f(x+\Delta \eta) d\eta
\end{equation}
where the second expression is in terms of $\eta$ defined by $\Delta \eta = \xi -x$. We assume that the signal $f$ has a globally convergent Taylor expansion so that for
normalised filter-kernels the application of the filter may be expressed as:
\begin{equation}
{\overline{f}}=f+\sum_{m=1}^{\infty} \Delta^{m}M_{m} f^{(m)}~~~;~~M_{m}=\frac{1}{m!} \int _{-\infty}^{\infty} G(\eta) \eta^{m} ~ d\eta
\end{equation}
where $f^{(m)}$ denotes the $m$-th derivative of $f$ and we introduced the {\bfi moments} $M_{m}$ of the filter, including a factor $1/m!$ for notational convenience. This expresses
the filtered signal in terms of derivatives of the unfiltered signal. Application to the turbulent stress tensor in one spatial dimension allows to write
\begin{eqnarray}
\tau&=& {\overline{u^{2}}}-{\overline{u}}^{2} \nonumber \\
&=& \Big\{ u^{2}+\sum_{m=1}^{\infty} \Delta^{m}M_{m} (u^{2})^{(m)} \Big\} - \Big\{ u+\sum_{m=1}^{\infty} \Delta^{m}M_{m} u^{(m)} \Big\}^{2} \nonumber \\
&=& \sum_{m=2}^{\infty} \Delta^{m}M_{m} \{ (u^{2})^{(m)} - 2 u u^{(m)} \} - \sum_{m=1}^{\infty}\sum_{n=1}^{\infty} \Delta^{m+n} M_{m}M_{n} u^{(m)} u^{(n)}
\label{tau_series}
\end{eqnarray}
where use was made of $(u^{2})'-2uu'=0$. This expression for the turbulent stress holds for general filters for which all moments $M_{m}$ exist. As an example, this includes
popular filters such as the top-hat, Helmholtz or Gaussian filter \cite{geurtsbook2003}, but it is equally valid for general compact support high-order filters
\cite{geurts_inverse}. Collecting terms according to their power of $\Delta$ we have after some rewriting
\begin{eqnarray}
\tau&=& \Delta^{2} (u')^{2} \Big\{ 2M_{2}-M_{1}^{2} \Big\} + \Delta^{3} \Big((u')^{2} \Big)' \Big\{ 3M_{3} -M_{1}M_{2} \Big\} \nonumber \\
&+& \Delta^{4} \Big\{ \Big((u')^{2}\Big)' \Big[ 4M_{4}-M_{1}M_{3}\Big] \! \! + \! \! (u'')^{2}\Big[2M_{1}M_{3}-M_{2}^{2}-2M_{4}\Big] \Big\} + \ldots
\label{tau_trunc}
\end{eqnarray}
This expression contains contributions from all terms $\sim \Delta^{n}$ with $n \geq 2$ and involves both even and odd order moments. Many popular filters are formulated in
terms of an even filter-kernel, i.e., $G(-\eta)=G(\eta)$. For these filters the odd moments are identically zero and correspondingly all contributions $\sim \Delta^{n}$ with $n$
odd, are zero as well. In case the filter is assumed to be an $N$-th order filter, i.e., by definition $M_{m}=\delta_{m0}$ for $m=0,1,\ldots, N-1$ at some $N \geq 1$, a
corresponding additional number of contributions of lower order in $\Delta$ is equal to zero. This provides some control over the formal magnitude of the turbulent stress tensor
and may be interpreted as a less strong filtering in the sense that the effective filter-width defined in (\ref{fildef}) decreases rapidly with increasing filter-order
\cite{bosgeurts_api1}.

For first - or second order filters that are commonly considered in large eddy simulation, the truncation of (\ref{tau_trunc}) at order $\Delta^{2}$ yields the standard
tensor-diffusivity model. To replace derivatives of $u$ in this expression with corresponding derivatives of ${\overline{u}}$ it is required to approximately invert the filter.
Consistent to the required order in $\Delta$, we may approximate the inversion of the filter simply as $u \approx {\overline{u}}$ which leads to
\begin{eqnarray}
m_{TD}&=& \Delta^{2} \{ 2M_{2}-M_{1}^{2} \} (\partial_{x}{\overline{u}})^{2}= \Delta^{2} V_{2} (\partial_{x}{\overline{u}})^{2} \nonumber \\
&=& \frac{1}{2}\{ 2M_{2}-M_{1}^{2} \}(k\Delta)^2 \Big( H^2(k\Delta)+ H^2(k\Delta) \cos(2kx) \Big)
\label{standard_td}
\end{eqnarray}
with corresponding flux
\begin{equation}
f_{TD}=k \Big(\{ 2M_{2}-M_{1}^{2} \}(k\Delta)^2 H^{2}(k\Delta) \Big) \sin(2kx)=kA_{TD}(k\Delta) \sin(2kx)
\label{tdflux}
\end{equation}
and, for notational convenience, we introduced the {\bfi filter variance} $V_{2}=2M_{2}-M_{1}^{2}$. In case a higher order accurate consistent truncation of (\ref{tau_trunc}) is
desired, or, when the filter itself is of higher order, then the expression of $u$ in terms of ${\overline{u}}$ and its derivatives becomes slightly more involved. However, this
poses no principal problems and one may invert the class of filters considered here at least formally up to arbitrary order. We will not consider these extensions but restrict
ourselves to the standard tensor-diffusivity model in (\ref{standard_td}).

The tensor-diffusivity model (\ref{standard_td}) is known to induce instabilities into a large eddy simulation \cite{vremantcfd,winckelmans}. These instabilities may be removed
by adding a (dynamic) eddy-viscosity contribution. However, the instabilities may also be addressed by turning to the spatially filtered tensor-diffusivity model instead, as was
observed in numerical experiments \cite{geurtsholm2002} and recently established analytically \cite{titi2004}. In fact, it was shown that spatial filtering of the
tensor-diffusivity model provides a closure that yields uniqueness and global regularity. For the filtered tensor-diffusivity model we find:
\begin{eqnarray}
m_{fTD}&=& \Delta^{2} V_{2} {\overline{(\partial_{x}{\overline{u}})^{2}}} \nonumber \\
&=& \frac{1}{2} V_{2}(k\Delta)^2 \Big( H^2(k\Delta)+ H^2(k\Delta)H(2k\Delta) \cos(2kx) \Big)
\end{eqnarray}
with corresponding flux
\begin{equation}
f_{fTD}=k \Big( V_{2}(k\Delta)^2 H^{2}(k\Delta)H(2k\Delta) \Big) \sin(2kx) =kA_{fTD}(k\Delta) \sin(2kx)
\label{ftd_flux}
\end{equation}
We notice the extra smoothing of the flux in (\ref{ftd_flux}) compared to (\ref{tdflux}), arising from the extra filter which adds a factor $H(2k\Delta)$. For all popular
large eddy filters this extra filter appears to damp the behaviour of the flux at high wavenumbers sufficiently to restore boundedness of the solution, however, at the expense of
increasing the computational costs.

In order to assess some of the properties of the regularisation and similarity subgrid models introduced above, we begin by considering the implications in case $k\Delta \ll 1$
and turn to properties at large $k\Delta$ momentarily. In the low wavenumber range the subgrid fluxes are small and we can obtain a precise impression of the type of deviations
that arise. By Taylor expansion of the characteristic amplitude functions we have:
\begin{eqnarray}
A_{\tau}(z)&=& -\Big(H''(0)-H'(0)^2\Big)z^2-\Big(H'''(0)-H'(0)H''(0) \Big)z^3 \nonumber \\
&-&\Big(\frac{7}{12}H''''(0)-\frac{1}{3}H'(0)H'''(0)-\frac{1}{4}H''(0)^2\Big)z^4+ \ldots \label{full_taylor_tau} \\
A_{L}(z) &=&-H'(0)z-\Big(\frac{3}{2}H''(0)+H'(0)^2\Big)z^2\nonumber \\
&-&\Big(\frac{7}{6}H'''(0)+2H'(0)H''(0)\Big)z^3\nonumber \\
&-&\Big(\frac{5}{8}H''''(0)+\frac{4}{3}H'(0)H'''(0)+\frac{3}{4}H''(0)^2\Big)z^4+\ldots \label{full_taylor_leray}
\end{eqnarray}
\begin{eqnarray}
A_{B}(z) &=& -\Big(H''(0)-H'(0)^2\Big) z^2-\Big(H'''(0)+H'(0)H''(0)-2H'(0)^3\Big)z^3 \nonumber \\
&-& \Big(\frac{7}{12} H''''(0)+\frac{5}{3} H'(0)H'''(0)+\frac{3}{4}H''(0)^2 \nonumber \\
&~&~~~~~~~~~~~~~~~-2H'(0)^2 H''(0)-H'(0)^4 \Big) z^4 + \ldots \\
A_{TD}(z)&=& V_{2}z^{2} + 2V_{2}H'(0)z^{3}+V_{2}\Big(  H'(0)^{2}+ H''(0)\Big) z^{4} + \ldots \\
A_{fTD}(z)&=&V_{2} z^2+4V_{2} H'(0) z^{3}+V_{2}\Big(5 H'(0)^{2}+3H''(0)\Big) z^{4} + \ldots 
\end{eqnarray}
where we assumed that all filters are normalised, i.e., $H(0)=1$. If we assume in addition that the filter-kernel $G$ is an even function, as is common in most large eddy
studies, then all odd order derivatives $H^{(2n+1)}(0)=0$ and the above expansions simplify to
\begin{eqnarray}
A_{\tau}(z)&=& -H''(0) z^2 -\Big(\frac{7}{12}H''''(0)-\frac{1}{4}H''(0)^2\Big)z^4+ \ldots \label{sym_taylor_tau}\\
A_{L}(z) &=&-\frac{3}{2}H''(0) z^2-\Big(\frac{5}{8}H''''(0)+\frac{3}{4}H''(0)^2\Big)z^4+\ldots \label{sym_taylor_leray} \\
A_{B}(z) &=& -H''(0) z^2- \Big(\frac{7}{12} H''''(0)+\frac{3}{4}H''(0)^2 \Big) z^4 + \ldots \\
A_{TD}(z)&=& V_{2}z^{2} +V_{2} H''(0) z^{4} + \ldots \\
A_{fTD}(z)&=&V_{2} z^2+3V_{2}H''(0) z^{4} + \ldots 
\end{eqnarray}

The expansions of the turbulent stress tensor in (\ref{full_taylor_tau}) or (\ref{sym_taylor_tau}) provide a point of reference for the different models. The Leray model gives
rise to a term $\sim z$ in (\ref{full_taylor_leray}), which is absent in the full turbulent stress tensor. Also, the second order contribution in (\ref{full_taylor_leray}) or
(\ref{sym_taylor_leray}) deviates systematically from the corresponding term in (\ref{full_taylor_tau}). Higher order corrections deviate as well but, since $7/12 \approx
5/8$, corrections involving the fourth order derivative of $H$ are still a fair approximation. These expansions establish that at low $k\Delta$ properties of the Leray (and
LANS$-\alpha$) model deviate markedly from properties of the exact closure problem. As mentioned before, this illustrates that the regularisation approach may be re-interpreted in
the language of the spatial filtering formulation for LES but it may not be directly derived from it.

The {\bfi Bardina similarity model} corresponds exactly at second order with the full turbulent stress tensor. Moreover, for a symmetric filter the fourth order corrections deviate from the exact turbulent stress in a way that is quite comparable to the Leray model. The expansions for the tensor-diffusivity model appear different at first sight. However, it is not hard to verify that $V_{2}=-(H''(0)-H'(0)^{2})$ and hence we notice that the lowest order terms for both the tensor-diffusivity and the filtered tensor-diffusivity model are exactly the same as in the full turbulent stress. Higher order corrections deviate in a way similar to what is observed in the Bardina model.

The truncated expansions are meaningful only for sufficiently low wavenumbers $k$. In this wavenumber range the subgrid flux is not particularly important for the dynamics of the
flow. To further assess the suitability of a subgrid closure for LES, the properties in the high wavenumber range are more important. We consider these next by explicitly
evaluating the dependence of the characteristic amplitude functions on $k\Delta$ for three well known symmetric LES filters, i.e., the top-hat ($th$), the Helmholtz ($H$) and the
Gaussian ($G$) filters. Specifically, the Fourier transform of these filters is given by
\begin{equation}
H_{th}(z)=\frac{\sin(z/2)}{z/2}~~;~~H_{H}(z)=\frac{1}{1+z^{2}/24}~~;~~H_{G}(z)=\exp(-\frac{z^{2}}{24})
\label{threefilters}
\end{equation}
The leading order expansion of each of these filters is given by $1-z^{2}/24+\ldots~$ . However, these filters differ essentially in their attenuation of high-$k$ solution
components. As $z \gg 1$ the amplitudes of these Fourier-transforms yield either a relatively slow algebraic decrease $|H_{th}|(z) \sim z^{-1}$ or $H_{H}(z) \sim z^{-2}$, or a
very fast exponential decrease for the Gaussian filter.

{
\begin{figure}[hbt]

\centering{
\psfig{figure=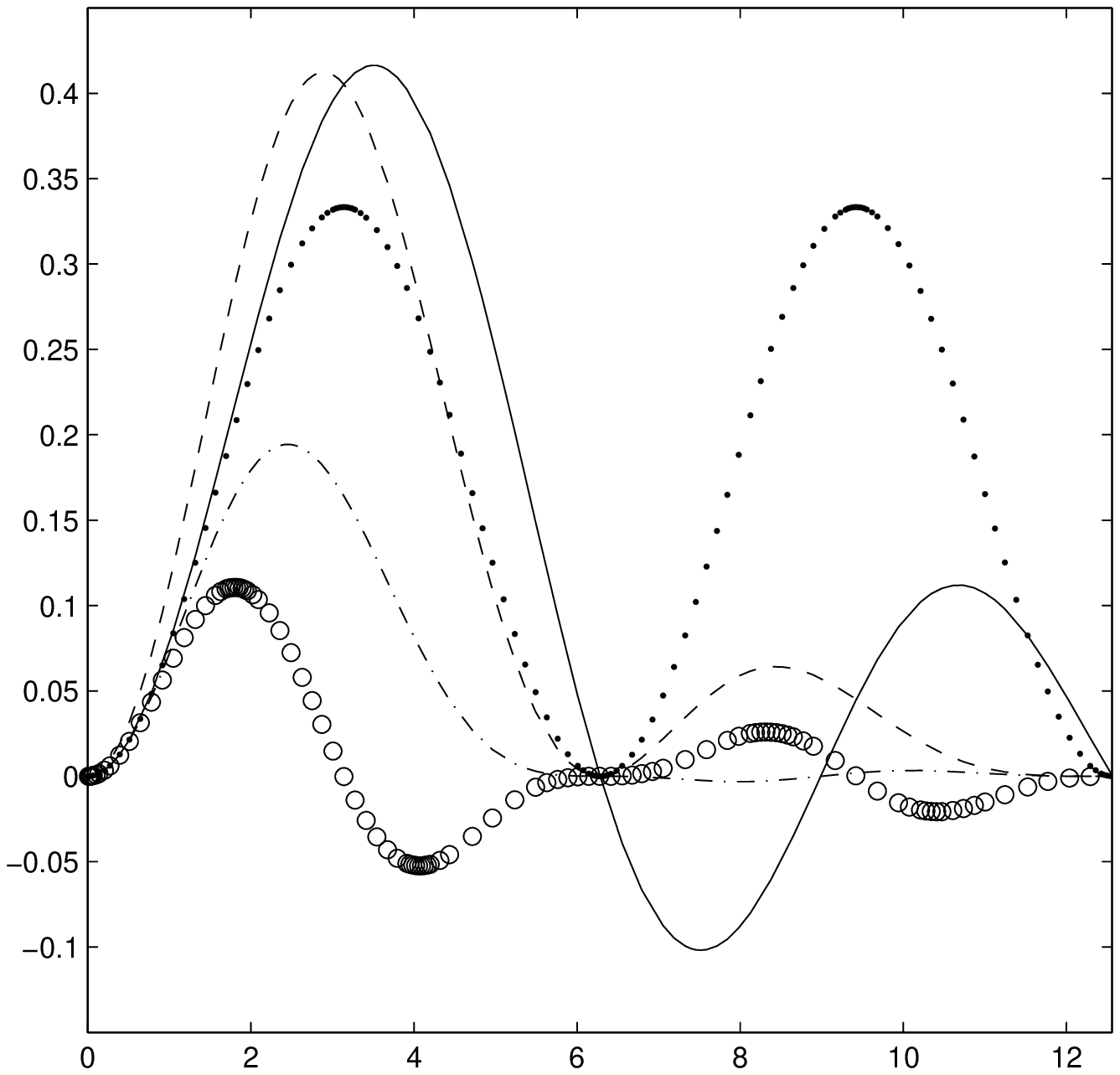,height=0.3\textwidth,width=0.28\textwidth}  (a)
\psfig{figure=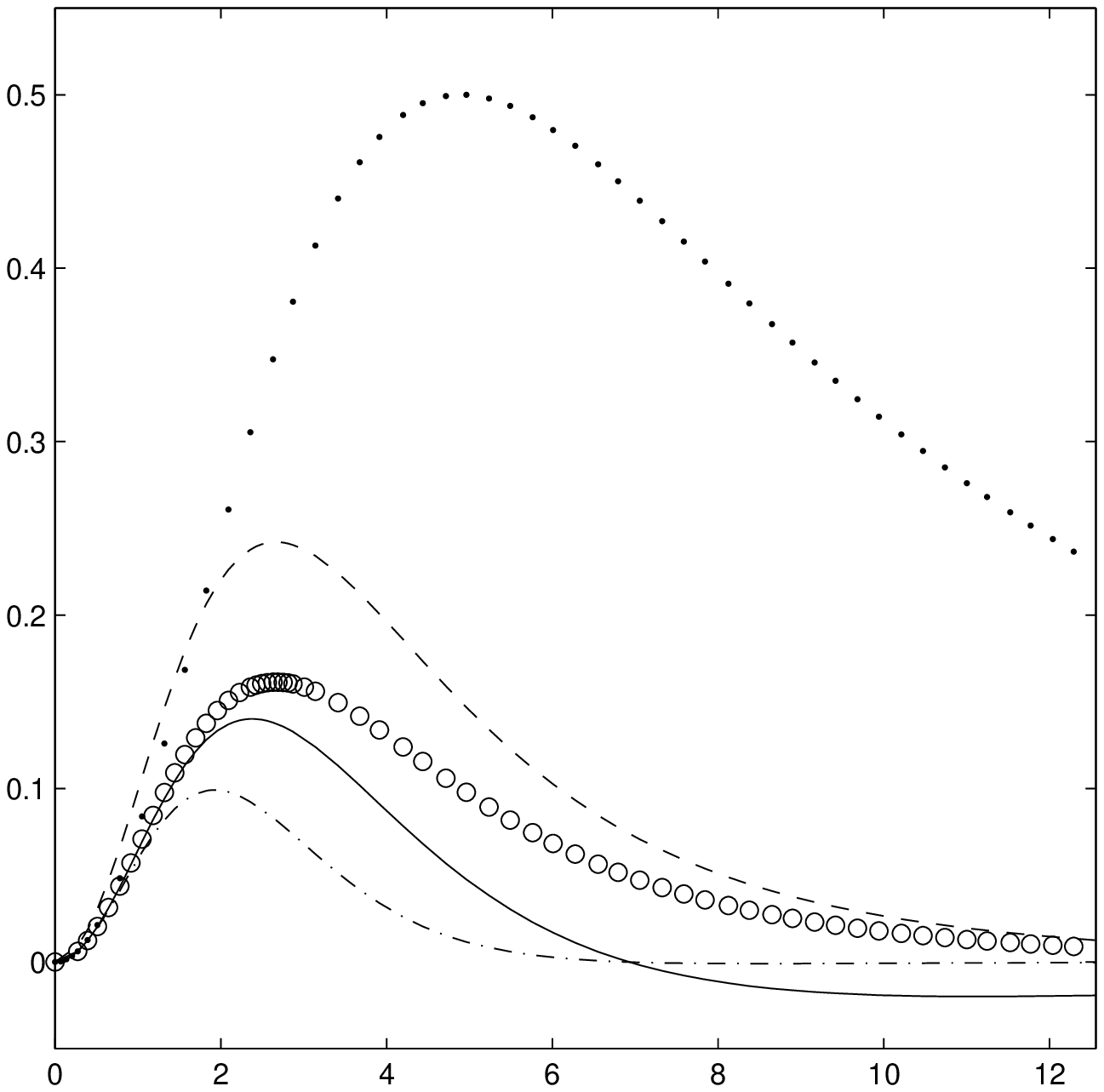,height=0.3\textwidth,width=0.28\textwidth} (b)
\psfig{figure=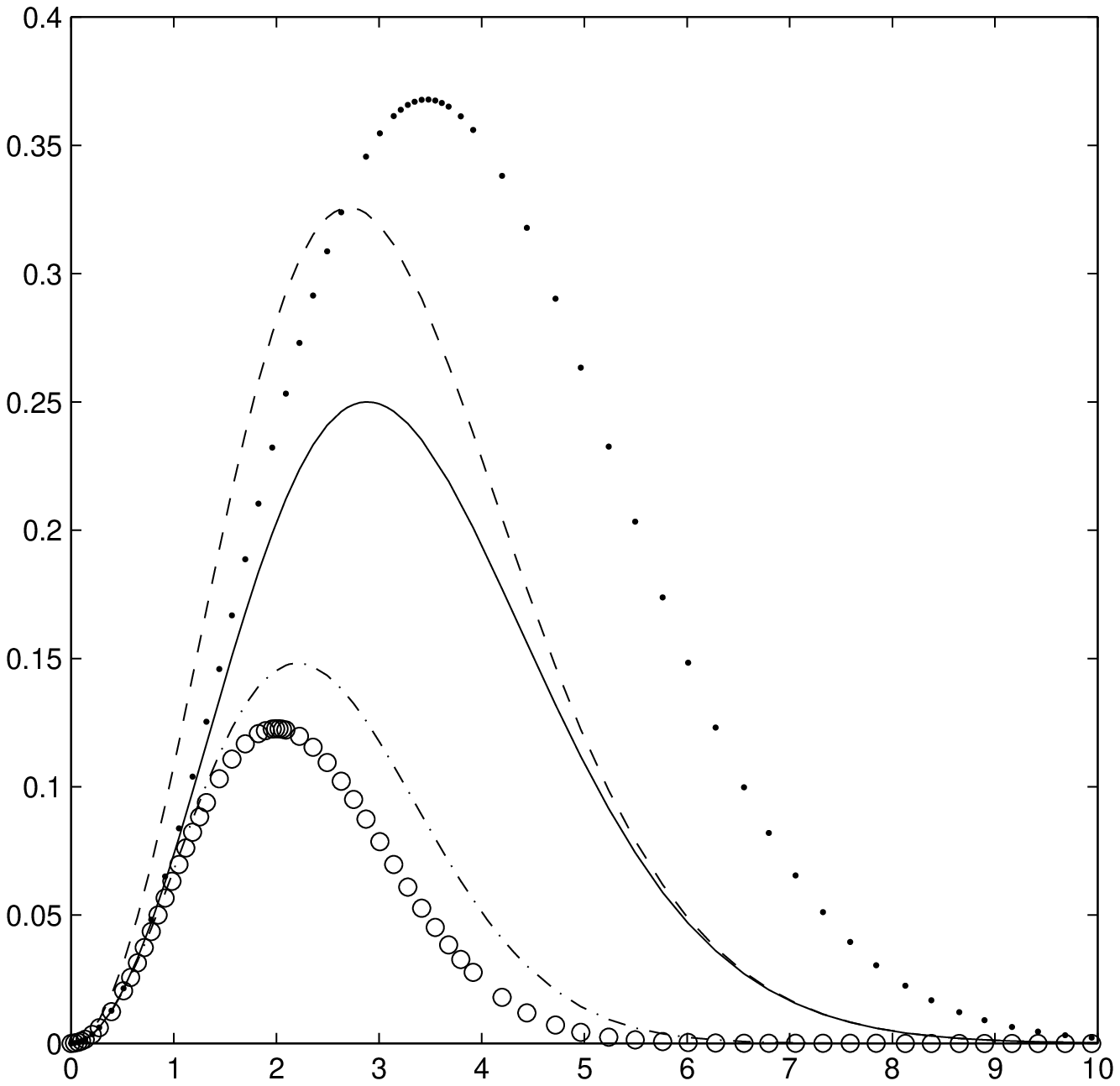,height=0.3\textwidth,width=0.28\textwidth} (c)
}

\caption{Characteristic amplitudes as function of {\mbox{$k\Delta$}} for the symmetric top-hat (a), Helmholtz (b) and Gaussian (c) filters: turbulent stress (solid), Leray
(dashed), Bardina (dash-dotted), tensor-diffusivity (dotted) and filtered tensor-diffusivity {\mbox{($\circ$)}}.}

\label{fourier_fig}
\end{figure}
}

The properties of the filter directly translate into the behaviour of the characteristic amplitude functions. Correspondingly, also the subgrid flux associated with the various
models is influenced. To illustrate this, we plotted the amplitude functions versus $k\Delta$ for the three filters (\ref{threefilters}) in figure~\ref{fourier_fig}. As point of
reference the characteristic amplitude of $\tau$ is shown as solid line, indicating either an oscillatory behaviour associated with the top-hat filter or a uni- or bi-modal
dependence on $k\Delta$ for the Gaussian and Helmholtz filters, respectively. The decay of the amplitude function for large $k\Delta$ differs in accordance with the asymptotic
behaviour of the Fourier transforms; the Gaussian filter results in the smallest `active' range of wavenumbers, while the top-hat filter provides the largest range.

The Leray model is seen to deviate already at quite low $k\Delta$, consistent with the Taylor expansion shown above. However, for the top-hat and Gaussian filters the
correspondence of the Leray model with the exact turbulent stress is better and over a wider range than that observed for the other models. For the top-hat filter the maximum
amplitude arises at slightly too low $k\Delta$, indicating that the Leray model may exaggerate the larger scales in a solution. In combination with the Gaussian filter the tail of
the amplitude function is nearly perfectly predicted by the Leray model, although there is a similar over-prediction of the maximum amplitude at slightly too low wavenumber. For
the Helmholtz filter the correspondence of the Leray model is still quite close but the filtered tensor-diffusivity model agrees somewhat better with the exact amplitude
function. None of the models properly captures the change of sign in the amplitude function of $\tau$ that arises in combination with the Helmholtz filter.

Turning to the Bardina model, we observe that the amplitude is strongly under-predicted and that the {\bfi spectral support} is too restricted, i.e., the values of $k\Delta$ for which
the amplitude is sufficiently different from zero is too small. The tensor-diffusivity model overestimates the amplitude in combination with the Helmholtz or the Gaussian filters
and shows a periodic dependence on $k\Delta$ in combination with the top-hat filter. In particular, this means that, in combination with the top-hat filter, the amplitude
function does not reduce to zero even for very small length-scales. These properties correlate directly with the underlying unstable nature of this model in actual simulations
\cite{vremantcfd,titi2004}. The filtered tensor-diffusivity model improves essentially on this behaviour but, apart from the combination with the Helmholtz filter, never seems to
achieve accurate levels of subgrid flux.

Although this single-mode analysis cannot be fully conclusive regarding the behaviour of any of these subgrid models in actual large eddy simulations, it does provide some
interesting first illustrations that will be seen to correlate well with the findings in turbulent mixing flow to which we turn in the next section.

\section{Dominant flow-features in turbulent mixing}
\label{picsmix}

In this section we first describe the turbulent mixing flow that is used in this paper to assess the quality of regularisation modelling. In subsection \ref{nummeth} we sketch
the numerical method and illustrate the global development of this canonical flow. Then, in subsection \ref{instant}, we present the reference large eddy simulation results
obtained with the Leray and LANS$-\alpha$ models, restricting ourselves to a discussion of the prominent dynamic flow-structures that arise and how well these are captured with
these models. A quantitative analysis is presented in sections \ref{leraymix} and \ref{nsalpha}.

\subsection{Numerical simulation of temporal mixing}
\label{nummeth}

The governing equations are discretise using the so-called method of lines. In the computations we consider the compressible formulation and perform simulations at a low
convective Mach number which was shown to provide essentially incompressible flow-dynamics \cite{vremanjfm}. We write the Navier-Stokes or LES equations concisely as
$\pr_{t}{\cal U}= {\cal F}({\cal U})$ where ${\cal U}$ denotes the state-vector containing,  e.g.,  (filtered) velocity and pressure, and ${\cal F}$ is the total flux, composed
of the convective, the viscous, and in case of LES also the subgrid fluxes.

The operator ${\cal F}$ contains first and second order partial derivatives with respect to the spatial coordinates.  The equations are discretised on a uniform rectangular grid
and the grid size in the $x_j$-direction is denoted by $h_j$. If we adopt a spatial discretization around a grid point ${\bf x}_{ijk}$, the operator ${\cal F}({\cal U})$ can be
approximated in a consistent manner by an algebraic expression $F_{ijk}(\{U_{\alpha \beta \gamma}\})$ where $\{ U_{\alpha \beta \gamma} \}$ denotes the state vectors in all the
grid-points that cover the domain. Usually, only neighbouring grid points around $(i,j,k)$ appear explicitly in $F_{ijk}$. After these steps we obtain a large system of ordinary
differential equations
\be
\frac{{\rm d}U_{ijk}(t)}{{\rm{dt}}}=F_{ijk}(\{U_{\alpha \beta \gamma} \})~~~;~~U_{ijk}(0)=U_{ijk}^{(0)}
\en
where $U_{ijk}(t)$ approximates ${\cal U}({\bf{x}}_{ijk},t)$ and $U_{ijk}^{(0)}$ represents the initial condition. Hence, in order to specify the numerical treatment, apart from
the initial and boundary conditions, the spatial discretization which gives rise to $F_{ijk}$ and the temporal integration need to be defined. We next introduce these separately
which is central to the method of lines.

The time stepping method which we adopt is an explicit four-stage compact-storage Runge-Kutta method \cite{geurtsbook2003}. When we consider the scalar differential equation
$du/dt=f(u)$, this Runge-Kutta method performs within one time step of size $\delta t$
\be
u^{(j)}=u^{(0)}+\beta_j \delta t f(u^{(j-1)})~~~(j=1,2,3,4)
\en
with
$u^{(0)}=u(t)$ and $u(t+\delta t)=u^{(4)}$. With the coefficients $\beta_1=1/4$, $\beta_2=1/3$, $\beta_3=1/2$ and $\beta_4=1$ this yields a second-order accurate time integration
method \cite{jameson}. The time step is determined by the stability restriction of the numerical scheme. It depends on the grid-size $h$ and the eigenvalues of the flux Jacobi
matrix of the numerical flux $f$. In a short-hand notation one may write $\delta t = {\rm CFL}~ h / |\lambda_{max}|$ where $|\lambda_{max}|$ denotes the eigenvalue of the flux
Jacobi matrix with maximal size, and ${\rm CFL}$ denotes the Courant-Friedrichs-Levy-number which depends on the specific choice of explicit time integration method. For the
present four-stage Runge-Kutta method a maximum CFL number of 2.4 can be established using a Von Neumann stability analysis. In the actual simulations we use ${\rm CFL}=1.5$,
which is suitable for both DNS and LES, irrespective of the specific subgrid model used.

In order to specify the spatial discretization we distinguish between the treatment of the convective and the viscous fluxes. We will only specify the  numerical approximation of
the $\pr_1$-operator; the $\pr_2$ and $\pr_3$-operators are treated analogously. Subgrid-terms are discretised with the same method as the viscous terms. Throughout, we will use
a second order finite volume method \cite{gk_fv} for both the viscous and the convective fluxes. The discretization of the convective terms is the cell vertex trapezoidal
rule, which is a weighted second-order central difference. In vertex $(i,j,k)$ the corresponding operator is denoted by $d_1$ and for the approximation of $\pr_{1}f$ it is
defined as
\bea
(d_1 f)_{i,j,k} &=& (s_{i+1,j,k}-s_{i-1,j,k})/(2h_1) \label{D1} \\
\mbox{with}~~~~ s_{i,j,k} & =& (g_{i,j-1,k}+2g_{i,j,k}+g_{i,j+1,k})/4
\nonumber \\
\mbox{and}~~~~ g_{i,j,k} & =& (f_{i,j,k-1}+2f_{i,j,k}+f_{i,j,k+1})/4
\nonumber
\ena
This illustrates the second-order central difference applied to a quantity $s_{ijk}$ which itself is a local average of $f$ over the $x_{2}$ and $x_{3}$ directions. The viscous
terms contain second-order derivatives which are treated by a consecutive application of two first order numerical derivatives. This requires for example that the gradient of the
velocity is calculated in centres of grid-cells. In centre $(i+\half,j+\half,k+\half)$ the corresponding discretization  $D_1 f$ has the form
\bea
(D_1 f)_{i+\half,j+\half,k+\half} &=& (s_{i+1,j+\half,k+\half} -s_{i,j+\half,k+\half})/h_1 \label{visc1} \\
\mbox{with}~~~~ s_{i,j+\half,k+\half} & =&
(f_{i,j,k}+ f_{i,j+1,k}+f_{i,j,k+1}+f_{i,j+1,k+1})/4
\nonumber
\ena
The second derivative is subsequently calculated in the point $(i,j,k)$ by applying the operator $D_{1}$ again, but now to the staggered approximations of the first derivative.
Specifically, we determine $(D_1 f)_{i,j,k}$ which, according to (\ref{visc1}), contains terms $s_{i+\half,j,k}$. These may be evaluated from the now known approximations at
$(i+\half,j+\half,k+\half)$ which are available after the first application of $D_1$ as defined in (\ref{visc1}). This basic numerical method, i.e., consisting of explicit second
order accurate four-stage Runge-Kutta time-stepping and second order finite volume discretization was adopted for the direct and large eddy simulations of a turbulent mixing flow to which we turn next.

\paragraph{Problem statement}
The three-dimensional temporal mixing layer is considered at a Reynolds number based on upper stream velocity and half the initial vorticity thickness of 50. This is sufficiently
high to allow a mixing transition to small scales. It is also sufficiently low to enable an accurate DNS that resolves all relevant turbulent scales on the computational mesh.
The governing equations are solved in a cubic geometry of side $\ell$ which is set equal to four times the wavelength of the most unstable mode according to linear stability
theory {($\ell=59$ in the present units)}. Periodic boundary conditions are imposed in the streamwise ($x_{1}$) and spanwise ($x_{3}$) direction, while in the normal ($x_{2}$)
direction the boundaries are free-slip walls. The initial condition is formed by mean profiles corresponding to constant pressure $p$, $u_{1}=\tanh(x_{2})$ for the streamwise velocity component and $u_{2}=u_{3}=0$. Superimposed on the mean profile are two- and three-dimensional perturbation modes obtained from linear stability theory. Further details of the problem statement for the three-dimensional temporal mixing layer may be found in \cite{vremanjfm}.

{



}

Visualisation of the DNS data demonstrates the roll-up of the fundamental linear instability and successive pairings. Four rollers with mainly negative spanwise vorticity emerge from the initial condition ($t=20$). After the first pairing ($t=40$)
the flow has become highly three-dimensional. Another pairing ($t=80$) yields a single roller in which the flow exhibits a complex structure with many regions of positive
spanwise vorticity.  This structure is a result of the transition to turbulence which has been triggered by the pairing process at $t=40$. The simulation is stopped at $t=100$,
since the single roller at $t=80$ does not undergo another pairing in this computational model. The accuracy of the simulation with $192^3$ cells is satisfactory for our purposes
as was demonstrated by comparing with computations using, e.g., a fourth order discretization, or coarser grids containing $64^3$ and $128^3$ cells \cite{vremanthesis}.

In the next subsection we will further specify the computational modelling for the large eddy simulation of this turbulent mixing flow and present some results that characterise
the global flow-features as predicted by the regularisation models. This provides a first, general assessment of the quality of these models.

\subsection{Capturing instantaneous mixing flow with regularised large eddy simulation}
\label{instant}

Throughout this paper, we consider large eddy simulations on three different grids with $32^{3}$, $64^{3}$ and $96^{3}$ cells, while keeping the filter-width fixed at
$\Delta=\ell/16$. This value of $\Delta$ provides a challenging test-case for large eddy simulation, studied first in \cite{vremanjfm}. Here, $\ell$ denotes the length of the
side of the cubical computational domain that was used. In this way the {\bfi subgrid resolution} $r=\Delta/h$ is varied explicitly, and we can separately influence the effects of numerical errors in the description. Moreover, we may assess independently the quality with which the LES models capture the physical features of the flow
by considering the predictions obtained for the (approximately) grid-independent situation that is obtained if $r$ is large enough \cite{gf2002}.

The possible numerical contamination of the smaller resolved scales was found to be well characterised by the subgrid resolution. In the simulations considered here $r$ varies between 2 and 6; for the lower value of $r$ the specific spatial discretization has some effect, even for mean flow properties, while for $r\approx 4-6$ a more acceptable
solution is obtained that is approximately grid-independent for most intends and purposes. The deviations from filtered DNS data that remain at $r=6$ are a direct measure of the deficiencies in the subgrid model only and would also be found in the same way for other spatial discretization methods.

The regularisation models require explicit filtering and (approximate) inversion. Moreover, the dynamic procedure that will be included for comparison purposes, requires explicit
(test-)filtering as well. Various, filters can be used. In this paper, we will adopt invertible numerical quadrature to approximate the top-hat filter. In this way we arrive at a
consistent representation of the top-hat filter and at the same time guarantee exact numerical inversion on the computational grid. In one dimension, symmetric numerical
convolution filtering $\ub=G*u$ corresponds to kernels
\begin{equation}
G(z)=\sum_{j=-m}^{m} a_{j} \delta(z-z_{j})~~;~~|z_{j}| \leq \Delta/2
\end{equation}
defined on $2m+1$ points. In particular, in this paper we consider the simplest three-point filters with $a_{0}=1-s$, $a_{1}=a_{-1}=s/2$ and $z_{0}=0$, $z_{1}=-z_{-1}=\Delta/2$.
Here we use $s=1/3$ which corresponds to Simpson quadrature of the top-hat filter. In actual simulations the resolved fields are known only on a set of  grid points
$\{x_{m}\}_{m=0}^{N}$. The application of the inverse $L^{-1}$ of this three-point filter can be specified using discrete Fourier transformation. In particular, for a general
discrete filtered solution $\{ \ub(x_{m})\}$ we have \cite{kgvg}
\be
L^{-1}(\ub_{m})={{\sum_{j=-n}^{n}}} \Big( \frac{s-1+\sqrt{1-2s}}{s} \Big)^{|j|} \frac{{{\ub_{m+rj/2}}}}{(1-2s)^{1/2}}
\en
where the subgrid resolution $r=\Delta/h$ is assumed to be even. An accurate and efficient inversion can be obtained with only a few terms, recovering the original signal to
within machine accuracy with $n \approx 10$. Filtering and inversion in three dimensions arises from composing three one dimensional filters.

A first qualitative test of subgrid models is obtained by studying the prediction of the dominant flow-structures in instantaneous solutions. So far in literature, failure of a
large eddy simulation to predict these global features of an evolving flow was not considered a decisive stumbling block for the particular subgrid model that was adopted.
Typically, it is argued that differences in an initial DNS-field concerning length-scales well below the filter-width $\Delta$ would not be of importance to the corresponding
filtered initial condition for LES. However, sensitive dependence of a turbulent flow to small-scale differences in initial conditions could result in significant effects on the
instantaneous solutions at later times. Hence, the correspondence between LES and DNS does not necessarily have to extend to instantaneous realizations of turbulent fields. This
point is certainly valid. However, at the modest Reynolds number considered here, and with the laminar initial condition that is used, one may expect that accurate subgrid models
will yield a strong correlation between filtered snapshots of the DNS solution and large eddy predictions. Correspondingly, a relevant and actually quite severe test-case for
LES, albeit at modest Reynolds numbers is obtained.

As a typical illustration of the mixing layer, the filtered DNS prediction of the vertical velocity and the corresponding Leray and LANS$-\alpha$ results can be compared in snapshots.We approximately eliminated the spatial discretization effects by using a resolution of $96^{3}$ in both large eddy simulations. We observe
that even at the instantaneous solution level, both the Leray and LANS$-\alpha$ models capture the `character' of the filtered solution. While the Leray solution appears to provide
a slight under-prediction of the influence of the small scales, the LANS$-\alpha$ solution corresponds more closely with the filtered DNS findings, restoring some of the small
scale variability. These instantaneous predictions are both much better than those obtained with the dynamic eddy-viscosity model which turn out to be too smooth.

{




}

The different regularisation models are known to have different effects on the tail of the resolved kinetic energy spectrum $E(k)$. In the Kolmogorov picture of homogeneous,
isotropic turbulence an inertial range in which $E(k) \sim k^{-5/3}$ develops over an extended range of wavenumbers $k$ up to a Kolmogorov wavenumber $k_{\eta} \sim 1/\eta$ where
$\eta$ is the viscous dissipation length-scale. This entire dynamic range needs to be properly captured in order to arrive at a reliable DNS. The Leray and LANS$-\alpha$ models
give rise to a spectrum in which there is a smooth transition from a $-5/3$ power law to a much steeper algebraic decay, beyond wavenumbers $\sim 1/\Delta$. The sharper decrease
of kinetic energy with wavenumber implies a corresponding strong reduction in required computational effort needed for the simulation of the relevant dynamic range. The
LANS$-\alpha$ model displays a tail of the spectrum $\sim k^{-3}$ while the Leray model decays even more steeply, as $\sim k^{-13/3}$ \cite{fht2001,ChHoOlTi2004}. The steeper decay using the Leray model is directly reflected in the smoother impression of instantaneous solutions. Hence, through the selection of $\Delta$ a direct external
control is achieved over the computational costs associated with the regularisation models. This is illustrated in figure~\ref{spectra}. In case an energy range of, say, $m$
decades is desired then all wavenumbers up to $k_{L}(m)$, $k_{\alpha}(m)$ and $k_{DNS}(m)$ need to be resolved for the Leray, LANS$-\alpha$ and DNS approaches respectively. This
corresponds to a significant difference in the associated computational expense, while all three simulations would provide excellent accuracy at least for all wavenumbers up to
$\sim 1/\Delta$.

{
\begin{figure}[hbt]

\centering{
\epsfig{figure=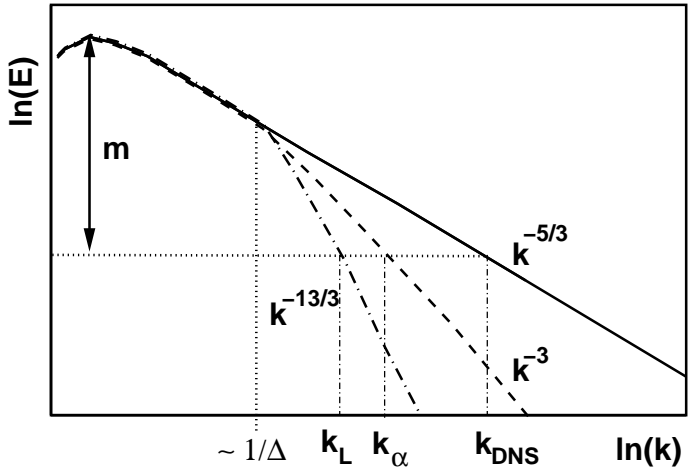,width=0.7\textwidth}
}
\caption{Sketch of resolved kinetic energy spectrum in a homogeneous, isotropic turbulence, displaying a $-5/3$ tail in DNS (solid), a $-3$ tail in LES using the LANS$-\alpha$ model
(dashed) and a $-13/3$ tail in LES using the Leray model (dash-dotted).}

\label{spectra}

\end{figure}
}

In the next section we will turn to a more quantitative assessment of the Leray model and investigate a range of flow features including mean flow as well as spectral properties.

\section{Leray predictions of turbulent mixing}
\label{leraymix}

In this section the Leray predictions for the turbulent mixing layer are discussed. First, in subsection~\ref{lowre} we restrict ourselves to a modest Reynolds number and compare
results directly with the available filtered DNS data. Subsequently, in subsection~\ref{highre} we turn to much higher Reynolds numbers and establish the robustness and
reliability of this computational model.

\subsection{Mixing at modest Reynolds numbers}
\label{lowre}

In order to assess the quality of a subgrid model it is important to address the accuracy of large eddy predictions for a variety of flow properties that characterise different
length-scale ranges. Moreover, a clear distinction should be made between the quality of the subgrid model and effects arising from the nonlinear interaction with discretization
errors at marginal spatial resolution \cite{gf2002,meyers2003}. We comply with these two general requirements by including LES predictions ranging from mean flow quantities and
statistics of fluctuations to spectra of turbulent kinetic energy, at different spatial resolutions. Moreover, we will consider the `inner operations' of the Leray model by
studying the contributions of the subgrid flux to the energy dynamics and establish the amount of forward and backward scatter.

\subsubsection{Mean flow properties.}

The resolved kinetic energy $E$ and the momentum thickness of the resolved flow $\delta$ are two characteristic mean flow properties of the temporal mixing layer. These
quantities are defined as
\begin{equation}
E=\int_{\Omega} \frac{1}{2} {\overline{u}}_{i}{\overline{u}}_{i}~d{\bf{x}}~~~;~~\delta=\frac14 \int_{-\ell/2}^{\ell/2} \Big(1-\langle {\overline{u}}_1\rangle (x_{2},t)\Big)
\Big(\langle {\overline{u}}_1 \rangle (x_{2},t) +1\Big)dx_2
\end{equation}
where $\Omega$ is the flow domain and $\langle \cdot \rangle$ denotes averaging over the homogeneous $x_{1}$ and $x_{3}$ directions. The evolution of $E$ illustrates the
transitional flow and subsequent self-similar decay in the turbulent regime. It depends primarily on the larger scales in the flow. Similarly, the momentum thickness is a
large-scale quantity that can be used as a measure for the progress of the mixing.

{
\begin{figure}[hbt]

\centerline{
\psfig{figure=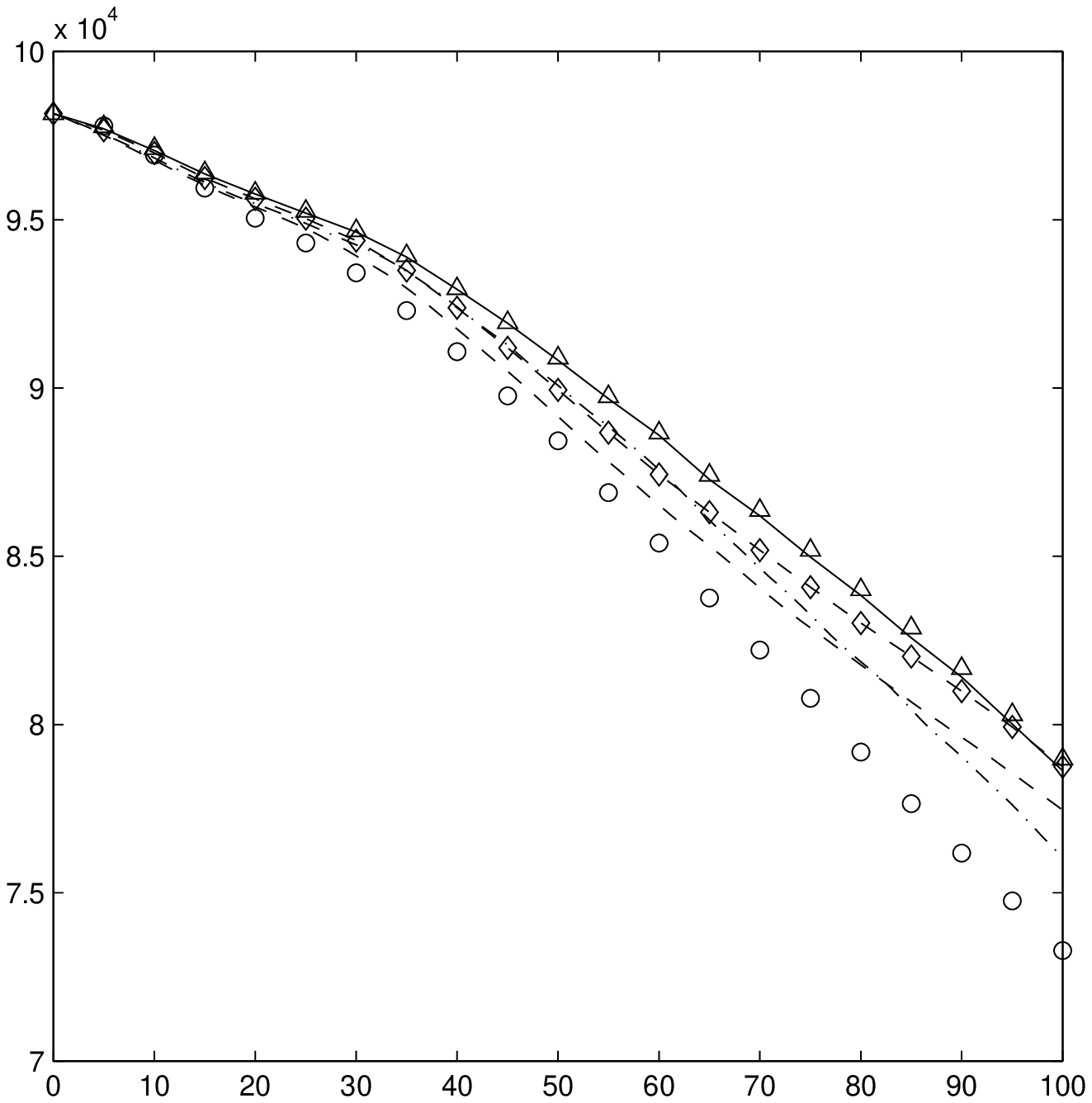,width=0.45\textwidth}(a)
\psfig{figure=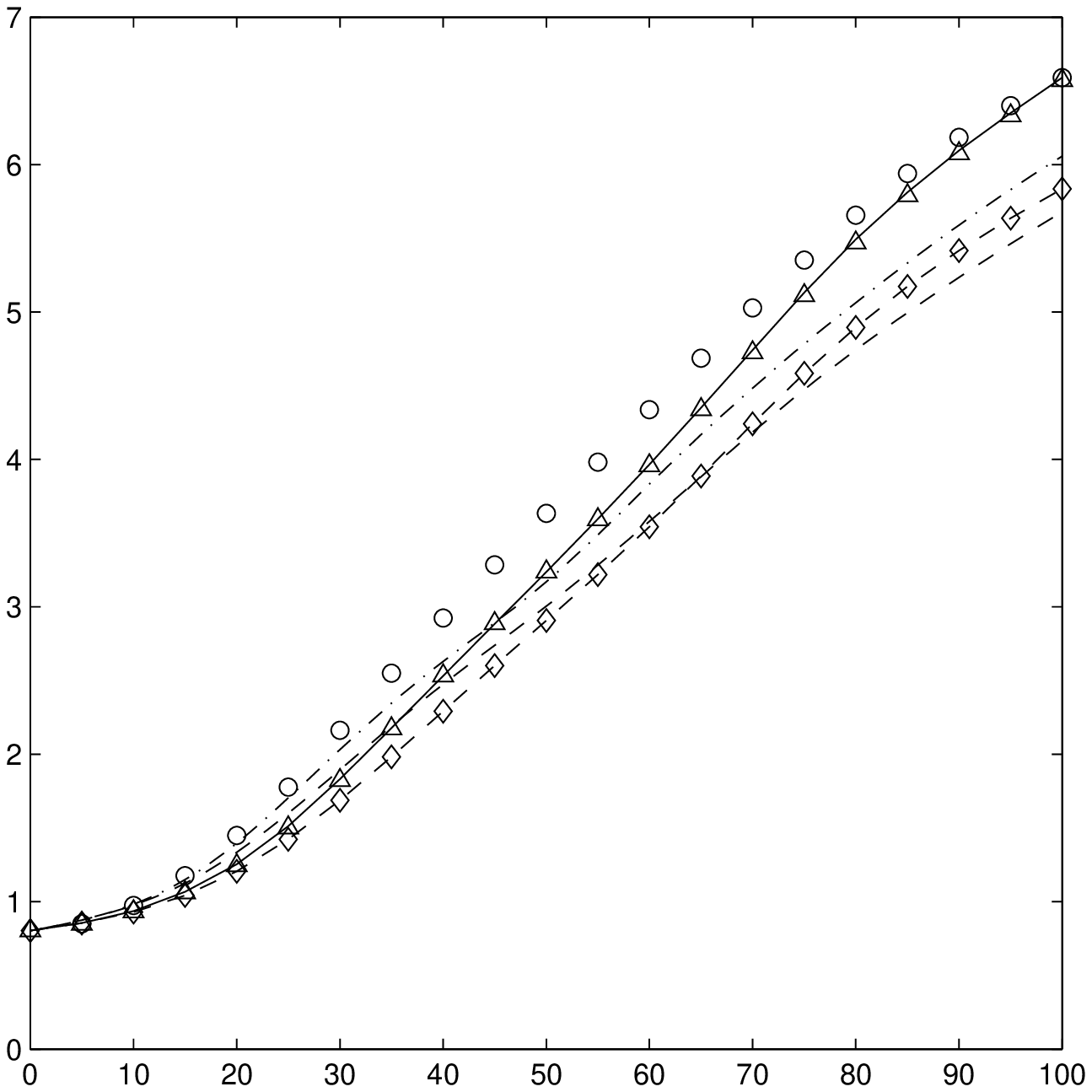,width=0.45\textwidth}(b)
}

\caption{Resolved kinetic energy $E$ (a) and momentum thickness $\delta$ (b): filtered DNS ($\circ$), Leray-model ($32^3$: dash-dotted, $64^3$: solid, $96^3$: $\triangle$),
dynamic model ($32^3$: dashed, $64^3$: dashed with  $\diamond$). A fixed filterwidth of $\ell/16$ was used at a Reynolds number $Re=50$.}

\label{ekin_momthi}

\end{figure}
}

The simulation results for $E$ and $\delta$ are collected in figure~\ref{ekin_momthi}. In an earlier LES study of this flow by Vreman {\it{et al.}} \cite{vremanjfm} it was
established that the dynamic eddy-viscosity model was among the subgrid models that performed best, compared to similarity - and other dynamic (mixed) models. We observe that
the kinetic energy $E$ is quite well represented during the laminar stages, but deviates considerably from the filtered DNS data during the turbulent stages. In particular, we
notice that this deviation becomes larger in case the subgrid resolution is increased at fixed filter-width. This indicates that the grid-independent solution corresponding to
the dynamic eddy-viscosity model does not provide sufficient dissipation on its own. Instead, the interaction with the particular spatial discretization method on coarser grids
is seen to provide additional decay of $E$. The capturing of the momentum thickness with the dynamic model is not very accurate in the important self-similar turbulent regime:
the grid-independent dynamic model yields a significant under-prediction of the mixing rate.

The Leray results indicate that the resolved kinetic energy is also overestimated in the turbulent regime. For the grid-independent solution the predicted decay rate $dE/dt$ is
nearly constant in the turbulent stages and corresponds well with the filtered DNS data. Moreover, the interaction with the spatial discretization method on coarse grids is seen
to result in an increased decay rate, similar to what was observed for the dynamic model. The Leray prediction for the momentum thickness compares significantly better with
filtered DNS results than was obtained with the dynamic model. We observe that some of the discrepancies between Leray and filtered DNS results are due to numerical contamination
on coarse grids. By increasing the resolution at fixed filterwidth $\Delta$, a good impression of the grid-independent solution to the modelled equations can be inferred using
$64^3$ -- $96^3$ grid-cells, i.e., $\Delta/h=4$ to $6$ \cite{gf2002}. Numerical contamination also plays a role in the dynamic model. The grid-independent prediction
for $\delta$ corresponding to the dynamic model appears less accurate than the corresponding Leray result.

\subsubsection{Statistics of velocity variations.}

{
\begin{figure}[hbt]

\centerline{
\psfig{figure=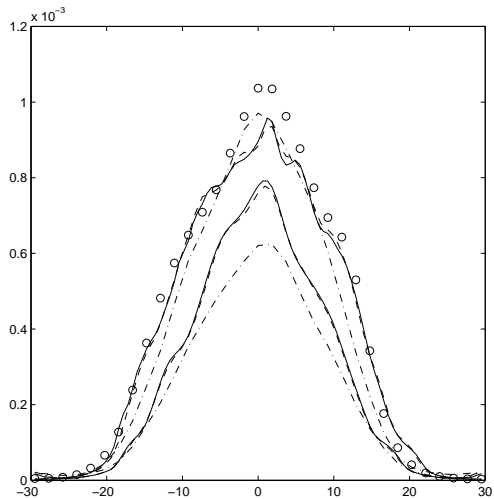,width=0.5\textwidth}
}

\caption{Convergence and comparison of the profile of the streamwise turbulent intensity $\langle v_1^2 \rangle^{1/2}$ at $t=80$ and $\Delta=\ell/16$. The filtered DNS data are
marked with $\circ$ and the convergence on $32^{3}$ (dash-dotted), $64^{3}$ (dashed) and $96^{3}$ (solid) is shown for the Leray model (top set of curves) and the dynamic model
(lower set of curves).}

\label{streamwise_leray_dynamic}

\end{figure}
}

In order to assess the quality of the Leray predictions for quantities that are more sensitive to small-scale information we next consider velocity variations defined by:
\begin{equation}
v_i={\overline{u}}_{i} - \langle {\overline{u}}_i \rangle
\end{equation}
where $\langle \cdot \rangle$ denotes averaging over the  homogeneous directions. Correlations among these velocity variation fields form the LES analogy of the Reynolds stresses
or turbulent intensities. A characteristic turbulent intensity $\langle v_{1}^{2} \rangle ^{1/2}$ is shown as a function of the normal coordinate $x_{2}$ in
figure~\ref{streamwise_leray_dynamic} in which we compare Leray and dynamic model predictions with filtered DNS data. We observe a clear convergence of the LES predictions with
increasing subgrid resolution $\Delta/h$. The turbulent flow associated with either of these subgrid models appears well resolved for $\Delta/h \gtrsim 4$ as observed above. The
approximately grid-independent predictions arising from the Leray and dynamic models are seen to slightly underestimate the streamwise turbulent intensity in the developed
turbulent regime. Compared with the dynamic model, the Leray predictions agree more closely with filtered DNS data. A similar level of agreement was found for the other turbulent
intensities $\langle v_{i}^{2} \rangle^{1/2}$ and the analogy of the Reynolds stress $\langle v_{1}v_{2}\rangle$.

\subsubsection{Resolved kinetic energy spectrum.}

A more detailed assessment is obtained from the streamwise kinetic energy spectrum shown in figure~\ref{spec_re50}. The dynamic model yields a significant under-prediction of the
intermediate and smaller retained scales, particularly for the approximately grid-independent solution. The Leray predictions are much better. On coarse grids, the interaction
with the spatial discretization method is seen to lead to a strong over-prediction of the smaller retained scales. However, at proper numerical subgrid resolution the situation
improves considerably and the Leray model is seen to capture all scales with high accuracy. A slight under-prediction of the smaller scales and a small over-prediction of the large
scales remains as systematic error in the grid-independent Leray solution.

{
\begin{figure}[hbt]

\centerline{
\psfig{figure=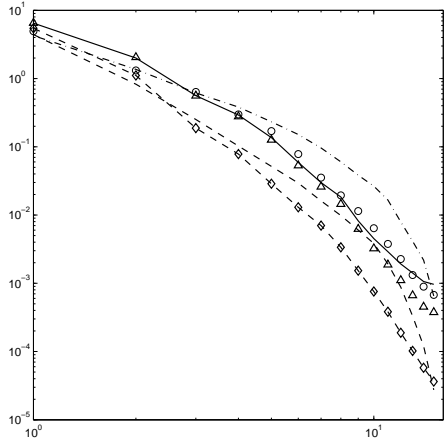,width=0.45\textwidth}
}

\caption{Streamwise kinetic energy spectrum $E$ at $t=75$: filtered DNS ($\circ$), Leray-model ($32^3$: dash-dotted, $64^3$: solid, $96^3$: $\triangle$), dynamic model ($32^3$:
dashed, $64^3$: dashed with $\diamond$). A fixed filterwidth of $\ell/16$ was used and the Reynolds number $Re=50$.}

\label{spec_re50}

\end{figure}
}

\subsubsection{Forward and backward scatter of energy.}

{
\begin{figure}[hbt]

\centerline{
\psfig{figure=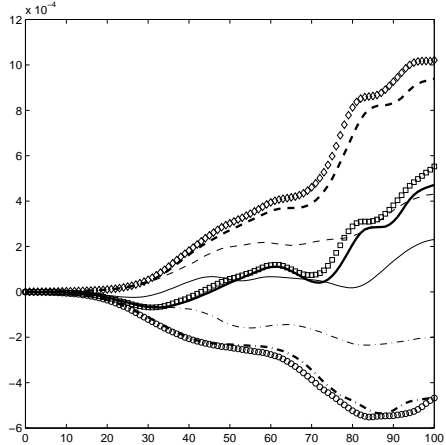,width=0.45\textwidth}
}

\caption{Total ($T_t$), forward ($T_f$), backward ($T_b$) energy transfer for the Leray model comparing three different grids. Thin lines correspond to $32^3$: $T_b$ dash-dotted,
$T_f$ dashed, $T_t$ solid, thick lines correspond to $64^3$: $T_b$ dash-dotted, $T_f$ dashed, $T_t$ solid and markers to $96^3$: $T_b$ $\circ$, $T_f$ $\diamond$, $T_t$ $\Box$. A
fixed filterwidth of $\ell/16$ was used and $Re=50$.}

\label{ekin_dyn_re_50}

\end{figure}
}

The `inner operations' of a subgrid model can be further characterised in terms of its dissipative or productive contributions to the evolution of the resolved kinetic energy
$E$. This evolution is governed by
\begin{equation}
\frac{dE}{dt}=\frac12 \int_{\Omega} \{ \frac{1}{Re}\ub_{i} \pr_{jj} \ub_{i} - \ub_{i} \pr_{j} \tau_{ij} \} ~ d{\bf{x}}= T_{v}-T_{t}
\end{equation}
where we introduced the viscous dissipation ($T_{v}$) and the total subgrid transfer ($T_{t}$). A positive total subgrid transfer $T_{t}$ contributes to a reduction of the
resolved kinetic energy and vice versa. The predicted kinetic energy evolution, corresponding to a given subgrid model may be found by replacing the turbulent stress tensor by
the assumed subgrid model.

To further analyse the dynamical behaviour, one may split the total subgrid contribution $T_{t}$ into a positive, i.e., forward scatter or dissipative, contribution $T_{f}$ and a
negative, i.e., backward scatter or reactive contribution $T_{b}$. To formalise this splitting, we introduce
\be
T_{f}=\frac14\int_{\Omega}(\ub_{i} \pr_{j} \tau_{ij} + |\ub_{i} \pr_{j} \tau_{ij}|)~d{\bf{x}}
~,~~
T_{b}=\frac14\int_{\Omega}(\ub_{i} \pr_{j} \tau_{ij} - |\ub_{i} \pr_{j} \tau_{ij}|)~d{\bf{x}}
\en
In figure~\ref{ekin_dyn_re_50} we collected the forward, backward and total energy transfer contributions. The individual contributions grow in magnitude in the transitional and
turbulent stages of the flow and are seen to be accurately captured numerically for $\Delta/h \gtrsim 4$. The Leray model is seen to contribute a significant backscatter
component to the dynamics of the flow. Also, the total transfer becomes strictly positive beyond the transitional stages $t \gtrsim 50$ which indicates that the Leray model on
average adds to the dissipation of kinetic energy.

Besides the performance of the Leray model at relatively low Reynolds number, the predictions in the asymptotic high Reynolds number range are important in case one is interested in applying this subgrid model to flow problems of realistic complexity. In the next subsection we consider this in some more detail.

\subsection{High Reynolds number limit}
\label{highre}

{
\begin{figure}[htb]

\centerline{
\psfig{figure=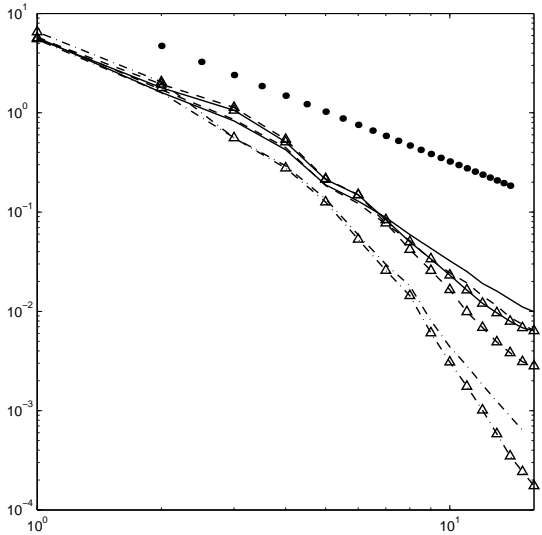,width=0.55\textwidth}
}

\caption{Streamwise kinetic energy spectrum at $t=75$ predicted by the Leray model: $Re=50$ ($64^3$: dash-dotted, $96^3$: dash-dotted, $\triangle$), $Re=500$ ($64^3$: dashed,
$96^3$: dashed, $\triangle$), $Re=5000$ ($64^{3}$: solid,$96^3$: solid, $\triangle$). A fixed filterwidth of $\ell/16$ was  used. The dotted line represents $k^{-5/3}$.}

\label{ekinspec_highre}

\end{figure}
}

A particularly appealing property of Leray modelling is its robustness at very high Reynolds numbers. This is illustrated in figure~\ref{ekinspec_highre} where we collected spectra of the resolved kinetic energy in the turbulent regime at three different Reynolds numbers. The lower $Re=50$ allows a corresponding direct numerical simulation. However, at $Re=500$ and $Re=5000$ such a direct simulation is not feasible and flow predictions in this regime are possible only using LES. The robustness at very high Reynolds numbers of the Leray model is quite unique for a subgrid model without an explicit eddy-viscosity contribution. Although comparison with filtered DNS data is impossible here, we observe that the smoothed Leray dynamics is essentially captured as $r=\Delta/h \geq 4$ \cite{gf2002}. The tail of the spectrum increases with $Re$, indicating a greater importance of small scale flow features. Improved subgrid resolution shows a reduction of these smallest scales, consistent with the reduced numerical error. At high $Re$ the spectrum corresponding to the Leray model tends to contain a region with approximately $k^{-5/3}$ behaviour, which is absent at $Re=50$.

{
\begin{figure}[htb]

\centerline{
\psfig{figure=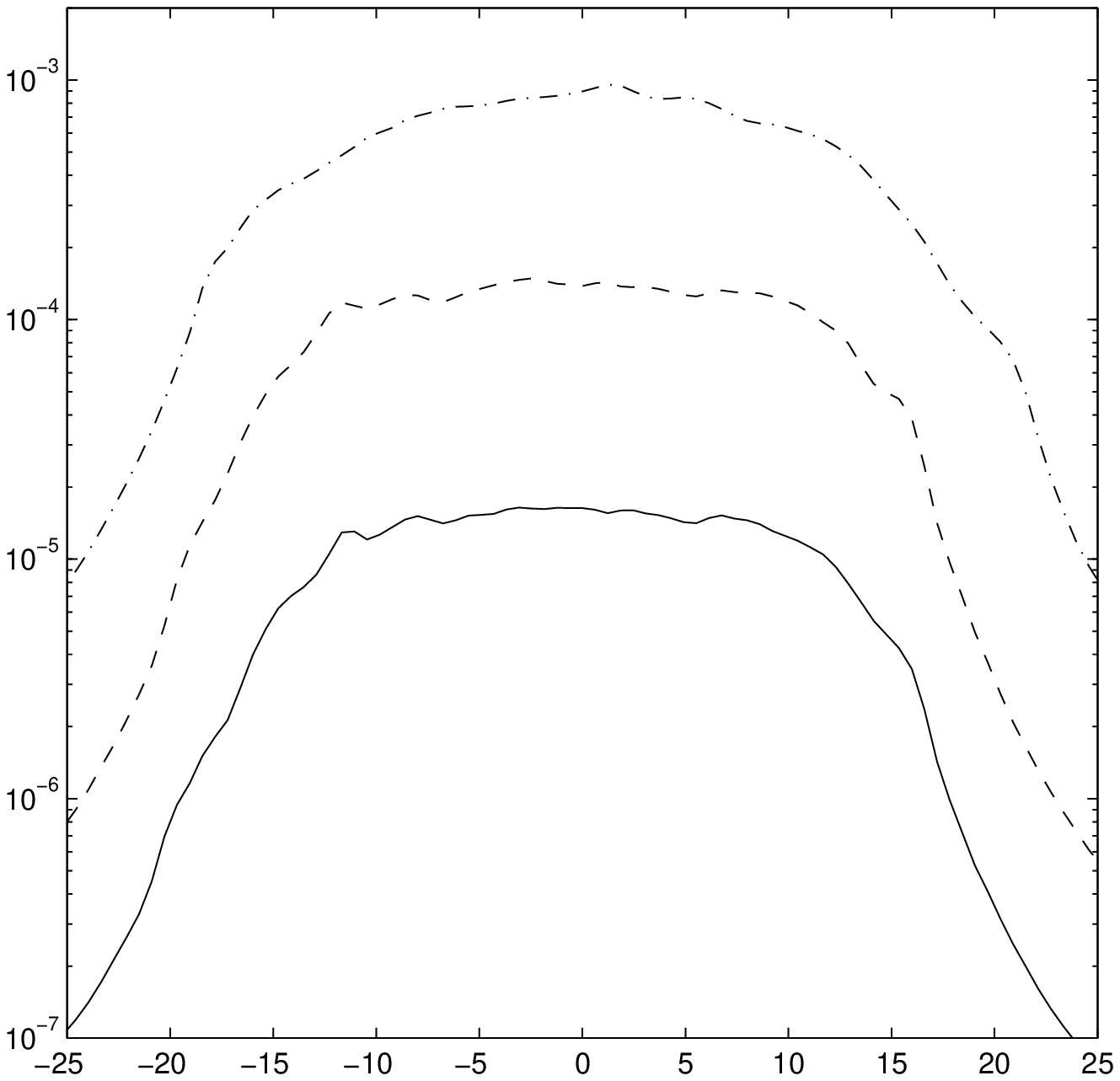,width=0.45\textwidth}(a)
\psfig{figure=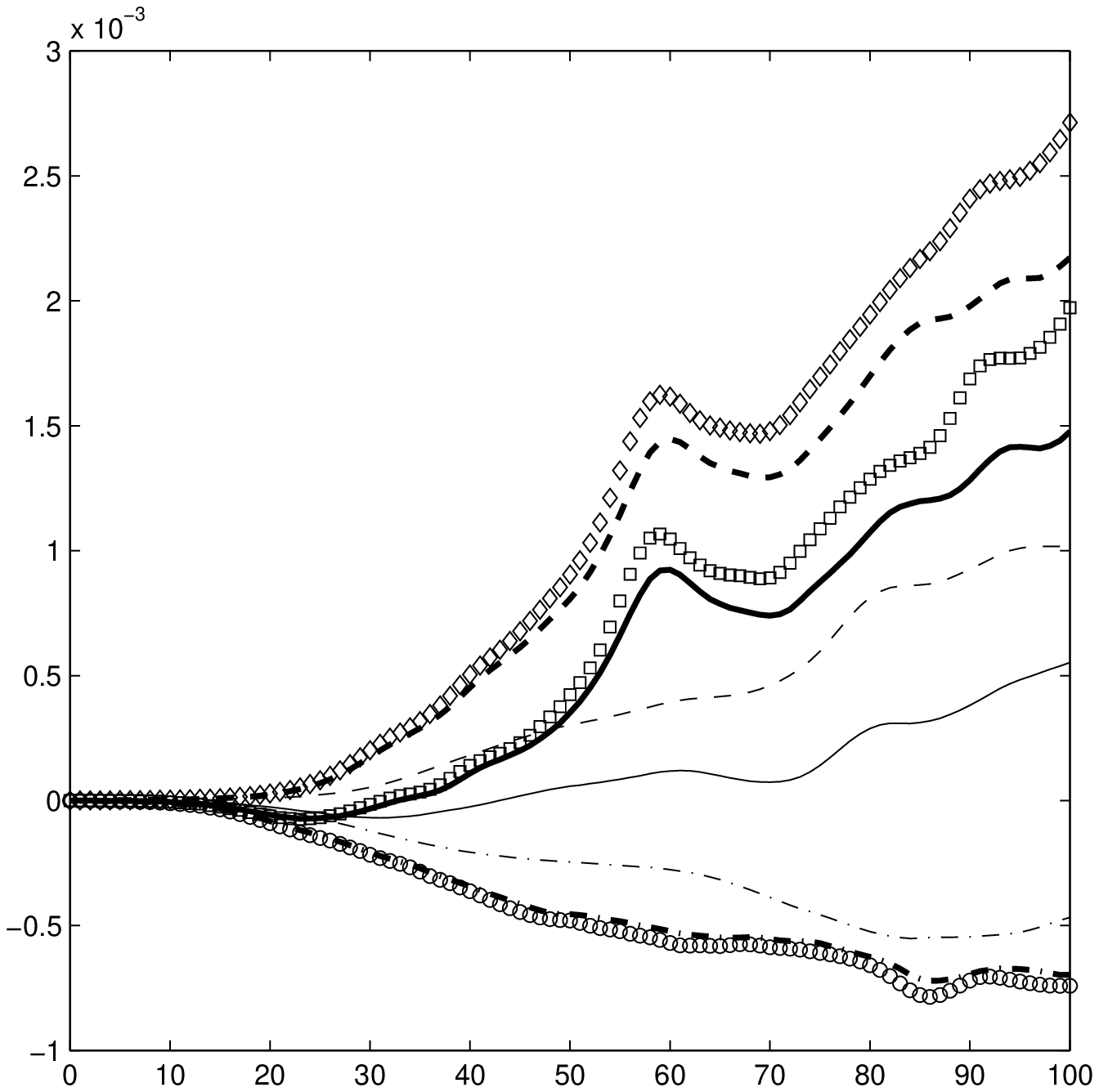,width=0.45\textwidth}(b)
}

\caption{The Reynolds dependence of the streamwise velocity fluctuations are shown in (a): $Re=50$ (dash-dotted), $Re=500$ (dashed) and $Re=5000$ (solid). In (b) the total
($T_t$), forward ($T_f$), backward ($T_b$) energy transfer for the Leray model are shown: thin lines correspond to $Re=50$: $T_b$ dash-dotted, $T_f$ dashed, $T_t$ solid, thick
lines correspond to $Re=500$: $T_b$ dash-dotted, $T_f$ dashed, $T_t$ solid and markers to $Re=5000$: $T_b$ $\circ$, $T_f$ $\diamond$, $T_t$ $\Box$. A fixed filterwidth of
$\ell/16$ and a grid with $96^3$ cells was used.}

\label{ekindyn_reystress_highre}

\end{figure}
}

The Leray prediction of the dependence of the streamwise velocity fluctuations on the Reynolds number is collected in figure~\ref{ekindyn_reystress_highre}(a). We observe that
$\langle v_{1}^{2}\rangle ^{1/2}$ decreases strongly with the Reynolds number, while the shape and width of the profiles remains quite similar for all Reynolds numbers. In this
range, at fixed filter-width $\Delta$, an increase in $Re$ by a factor of 10 results in a similar reduction of the maximum of the streamwise velocity fluctuations. The subgrid
transfer of resolved kinetic energy is illustrated in figure~\ref{ekindyn_reystress_highre}(b). We observe that an increase in $Re$ gives rise to an increase in the magnitude of
each of the distinguished components to the evolution of $E$. The total transfer is positive in the developed stages, thereby contributing to an increased dissipation of energy
as $Re$ increases. The backscatter $T_b$ appears to approach an asymptotic limit while the forward scatter $T_{f}$ is still strongly increasing with $Re$.

As was described in subsection~\ref{regclose}, the Leray model is part of the LANS$-\alpha$ regularisation. The LANS$-\alpha$ approach provides consistency of the large eddy
simulation with the filtered Kelvin circulation theorem. In the next section we will investigate what consequences this important extension of the Leray model has for the LES
capturing of turbulent mixing.

\section{LANS$-\alpha$ improvements and limitations}
\label{nsalpha}

In this section we will illustrate some improvements over Leray and dynamic model predictions that arise when the LANS$-\alpha$ subgrid model is adopted. As explained in subsection~\ref{nummeth}, the LANS$-\alpha$ model appears to better represent the small-scale variability that is contained in the turbulent flow and to correspond quite closely with filtered DNS data. This suggests also that derived macroscopic flow properties such as the resolved kinetic energy or the momentum thickness are better described using the LANS$-\alpha$ model. In subsection~\ref{alphamixing} we will investigate a number of flow properties to quantify these improvements. A practical consequence of retaining more small-scale variability in the numerical solution is that the required resolution also needs to be increased. A discussion of these computational aspects and corresponding practical limitations for the LANS$-\alpha$ model is presented in subsection~\ref{alpharesolution}.

\subsection{LANS$-\alpha$ prediction of turbulent mixing.}
\label{alphamixing}

In order to discuss the alterations in LES predictions that arise from adopting the LANS$-\alpha$ subgrid model, it is essential to distinguish between the findings associated with
the approximately grid-independent solution and the convergence process toward this solution. This distinction should of course be made for every subgrid model but it is all the
more relevant for the LANS$-\alpha$ parameterisation since this subgrid model allows the existence of a significant level of small-scale variability in the resolved flow. The LANS$-\alpha$ model corresponds very closely to filtered DNS snapshots, provided the subgrid resolution is adequate.
Under-resolution of the LANS$-\alpha$ model will constitute a source for strong numerical contamination which is more characteristic of the spatial discretization that was used than
a measure for the quality of the subgrid model. Therefore, in this subsection we will first turn to the prediction of the kinetic energy and momentum thickness obtained at high
subgrid resolution and consider the convergence process afterwards.

{
\begin{figure}[htb]

\centerline{
\psfig{figure=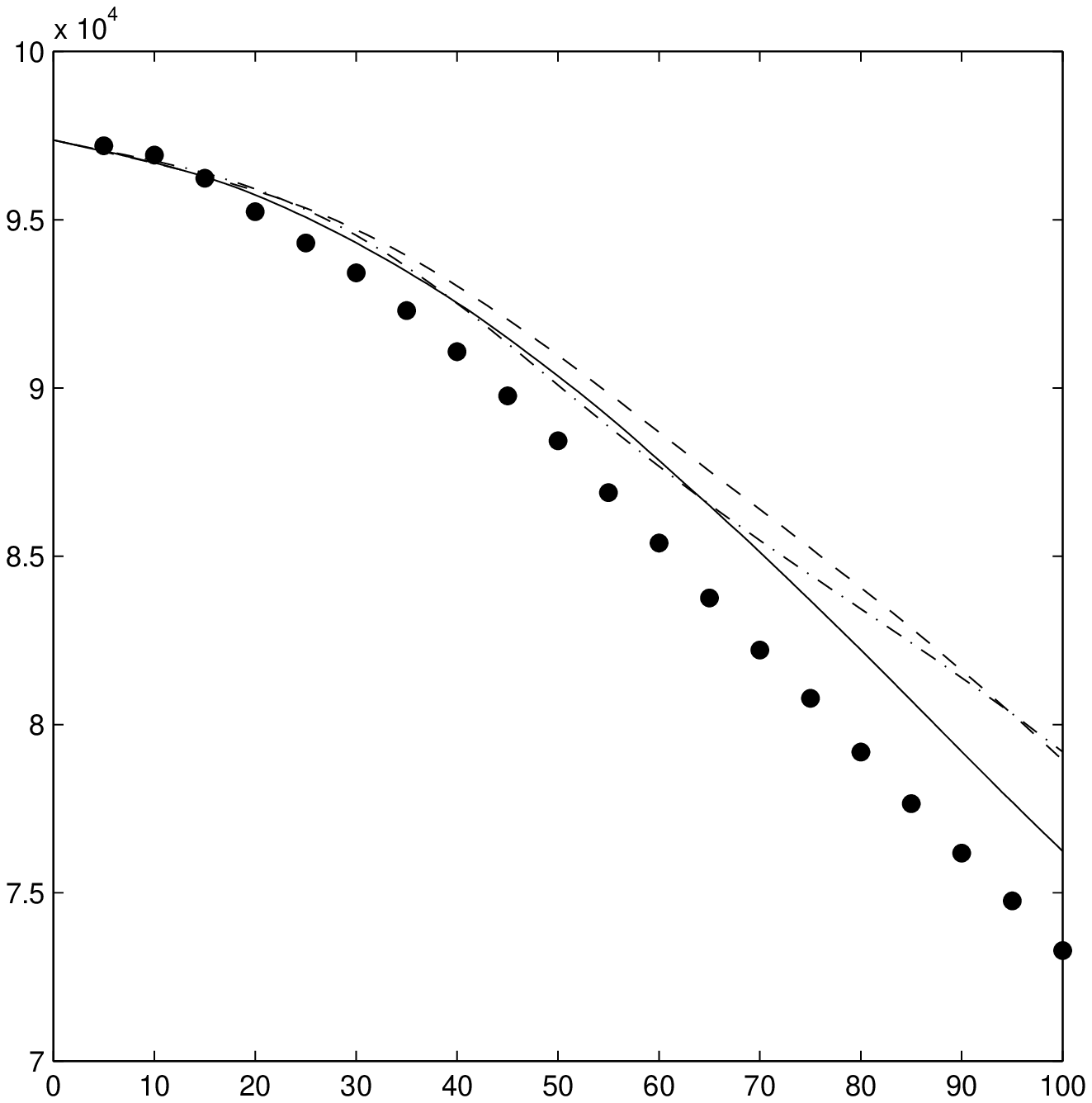,width=0.45\textwidth}(a)
\psfig{figure=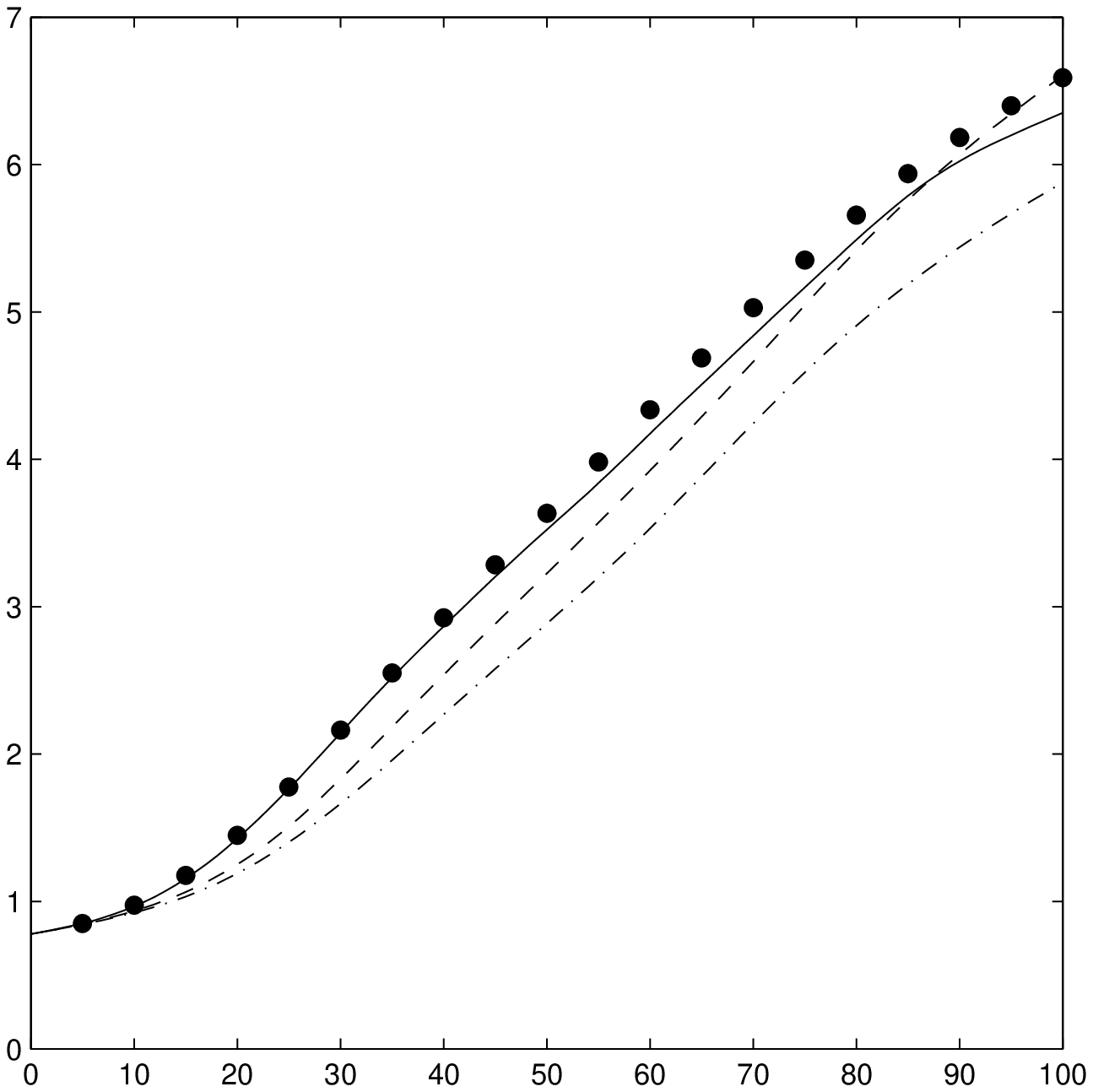,width=0.45\textwidth}(b)
}

\caption{Resolved kinetic energy $E$ (a) and momentum thickness $\delta$ (b) at $Re=50$ for $\Delta=\ell/16$ and three LES models: LANS$-\alpha$ (solid), Leray (dashed) and
dynamic model (dash-dotted), compared with filtered DNS (solid circles) corresponding to the approximately grid-independent prediction at $96^{3}$.}

\label{figmomthi}

\end{figure}
}

The evolution of the resolved kinetic energy and the momentum thickness is shown in figure~\ref{figmomthi}. In this figure the approximately grid-independent predictions obtained
at a resolution $96^3$, i.e., $r=6$ are collected. As was established in the previous section, we observe a strong improvement of the prediction of $\delta$ using the Leray model
compared to the dynamic model, whereas the resolved kinetic energy is predicted at a similar level of accuracy. In addition, the LANS$-\alpha$ results are seen to constitute a
further significant improvement for $\delta$ and agree almost perfectly with the filtered DNS results. The resolved kinetic energy is also seen to be much better described by the
LANS$-\alpha$ model; although the level of kinetic energy is slightly over-predicted, the decay rate $dE/dt$ is very close to that seen in the filtered DNS data. For both $E$ and
$\delta$ the improvement is remarkable in the early laminar and transitional regime, which appears to set the stage for the accurate prediction of $\delta$ in the developed
turbulent flow. Only in the very late stages of the simulation is a deviation seen to develop in the prediction for $\delta$, and this may be due to a less than complete resolution of the smallest retained scales. These simulation results clearly illustrate the improvements in the description of the physics of the flow by the LANS$-\alpha$ model which become available in case the spatial resolution is adequate.

{
\begin{figure}[hbt]

\centerline{
\psfig{figure=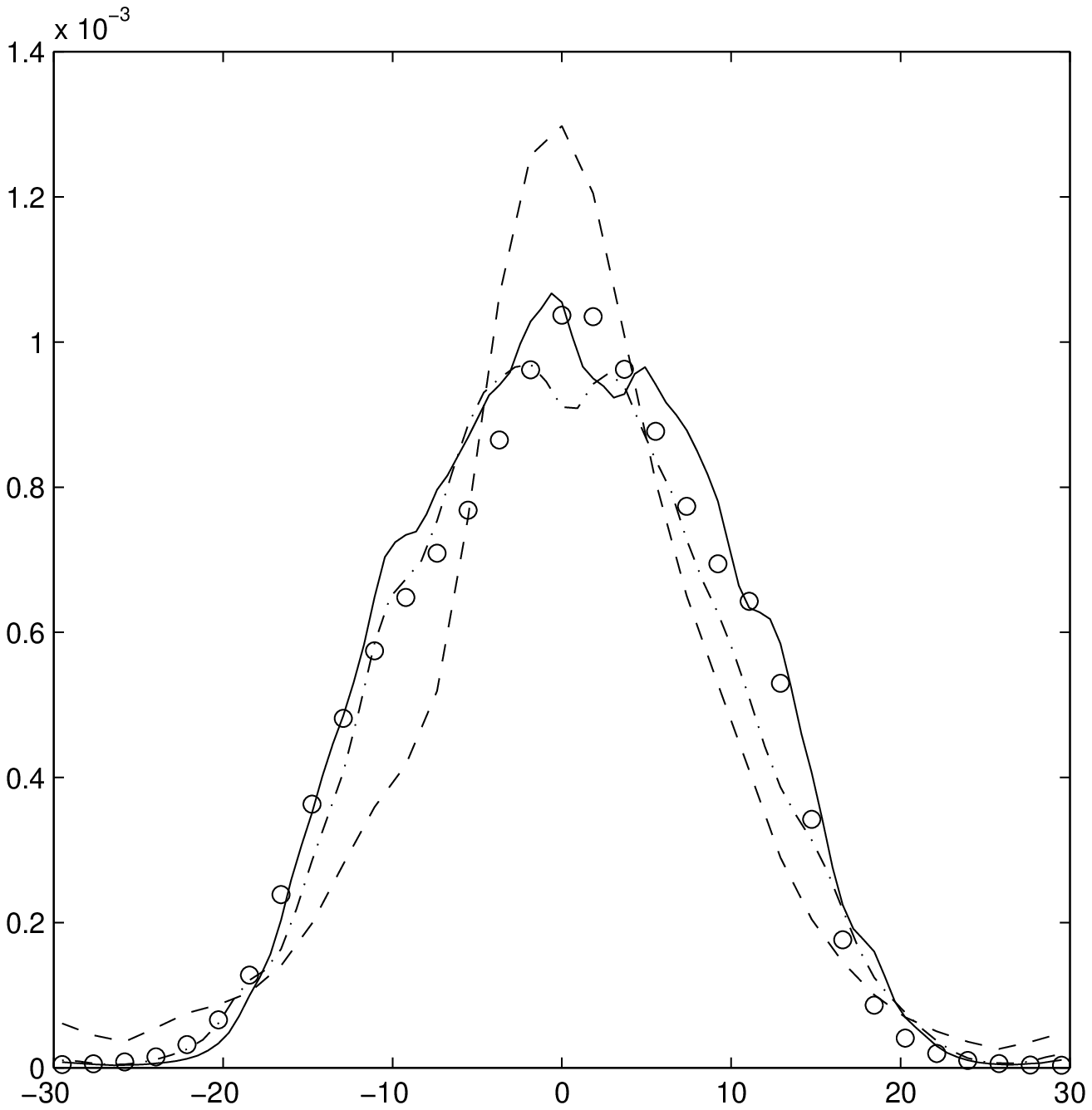,width=0.45\textwidth}(a)
\psfig{figure=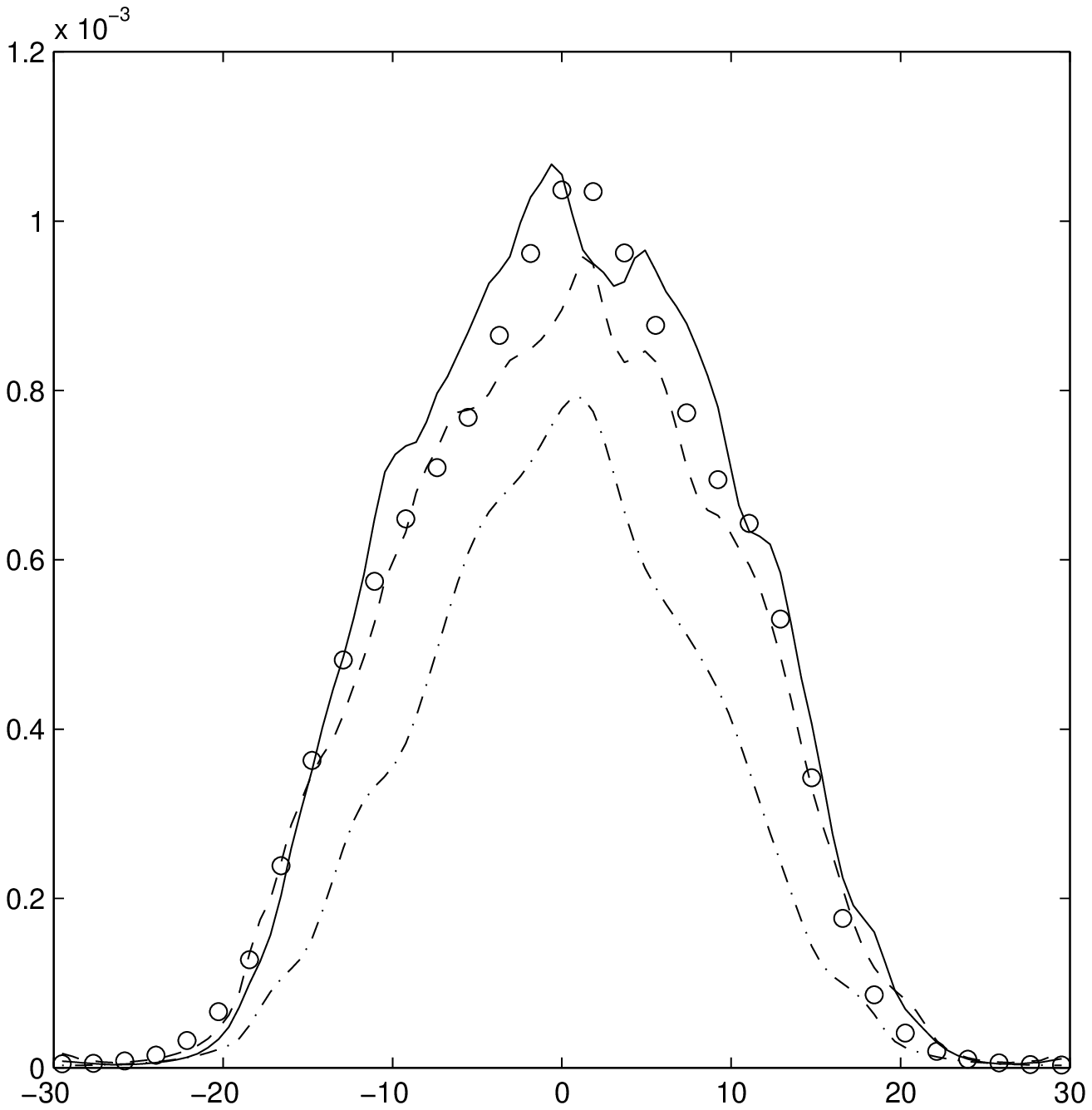,width=0.45\textwidth}(b)
}

\caption{Convergence and comparison of the profile of the streamwise turbulent intensity $\langle v_1^2 \rangle^{1/2}$ at $t=80$ and with $\Delta=\ell/16$. The filtered DNS data
are marked with $\circ$. The convergence of the LANS$-\alpha$ predictions on $32^{3}$ (dashed), $64^{3}$ (dash-dotted) and $96^{3}$ (solid) is shown in (a) and the approximately
grid-independent results are shown in (b) for the Leray model (dashed), the dynamic model (dash-dotted) and the LANS$-\alpha$ model (solid).}

\label{streamwise_nsalpha_improved}

\end{figure}
}

The prediction of the streamwise turbulent intensity is shown in figure~\ref{streamwise_nsalpha_improved}. We notice that under-resolution of the LANS$-\alpha$ model results in too
high levels of turbulent intensity. Further visualisation of the LANS$-\alpha$ flow at a resolution of $32^{3}$ showed snapshots which are dominated by an unphysically high level
of small scale features. However, at appropriate spatial resolution we observe a fair approximation of a grid-independent solution that is seen to correspond very well with the
filtered DNS data in figure~\ref{streamwise_nsalpha_improved}(b). The LANS$-\alpha$ and Leray predictions are both accurate representations of the filtered DNS findings with a
slight improvement of the predicted maximal turbulent intensities near the centerline due to the LANS$-\alpha$ model. There are certainly considerable improvements over the results
that are obtained when using the dynamic model. To put this further into perspective, an earlier study by Vreman {\it{et al.}} \cite{vremanjfm} established that the dynamic
model was among the more accurate models compared to the Bardina similarity model, the nonlinear or gradient model and other dynamic (mixed) models.

{
\begin{figure}[hbt]

\centerline{
\psfig{figure=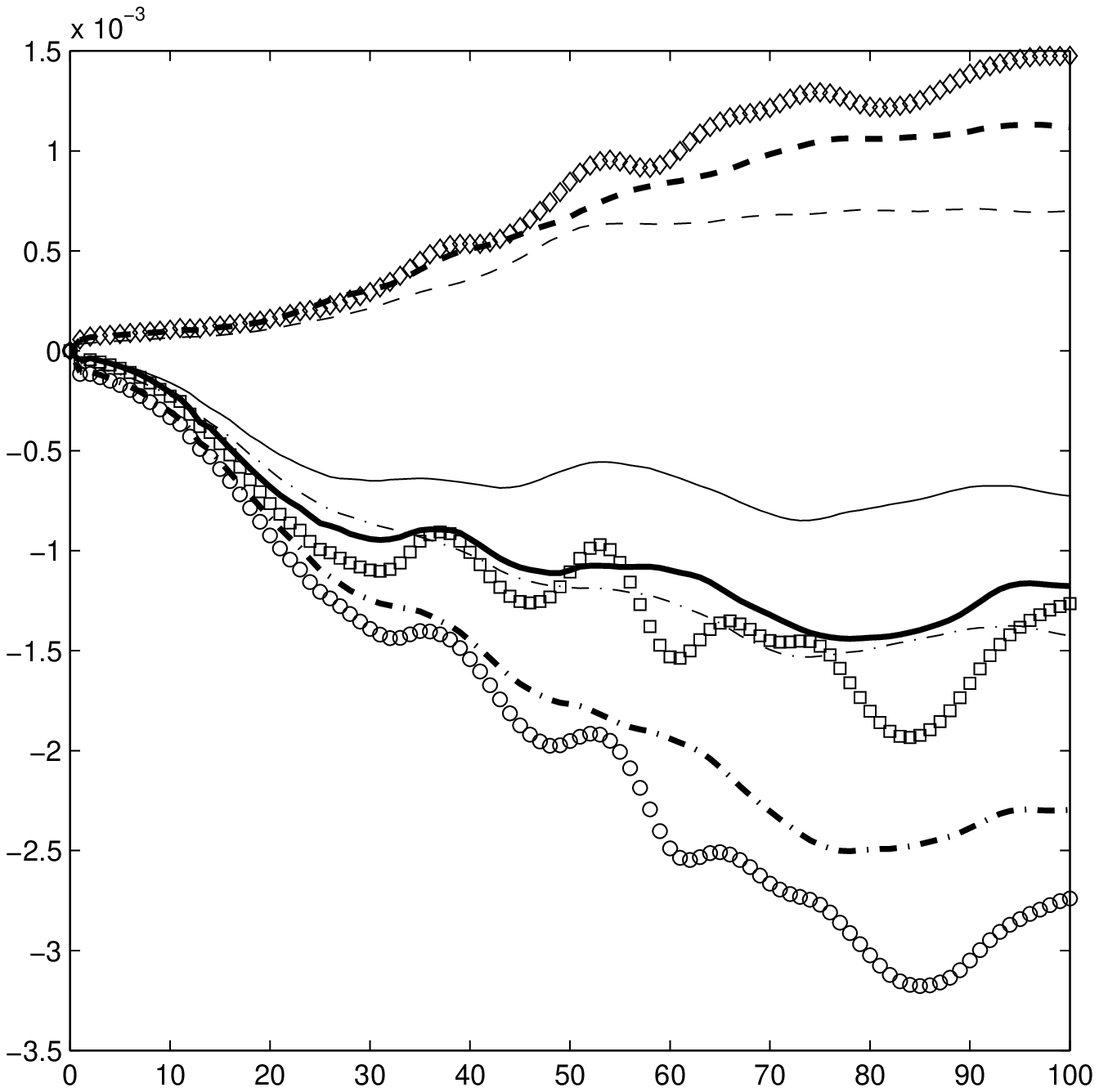,width=0.45\textwidth}(a)
\psfig{figure=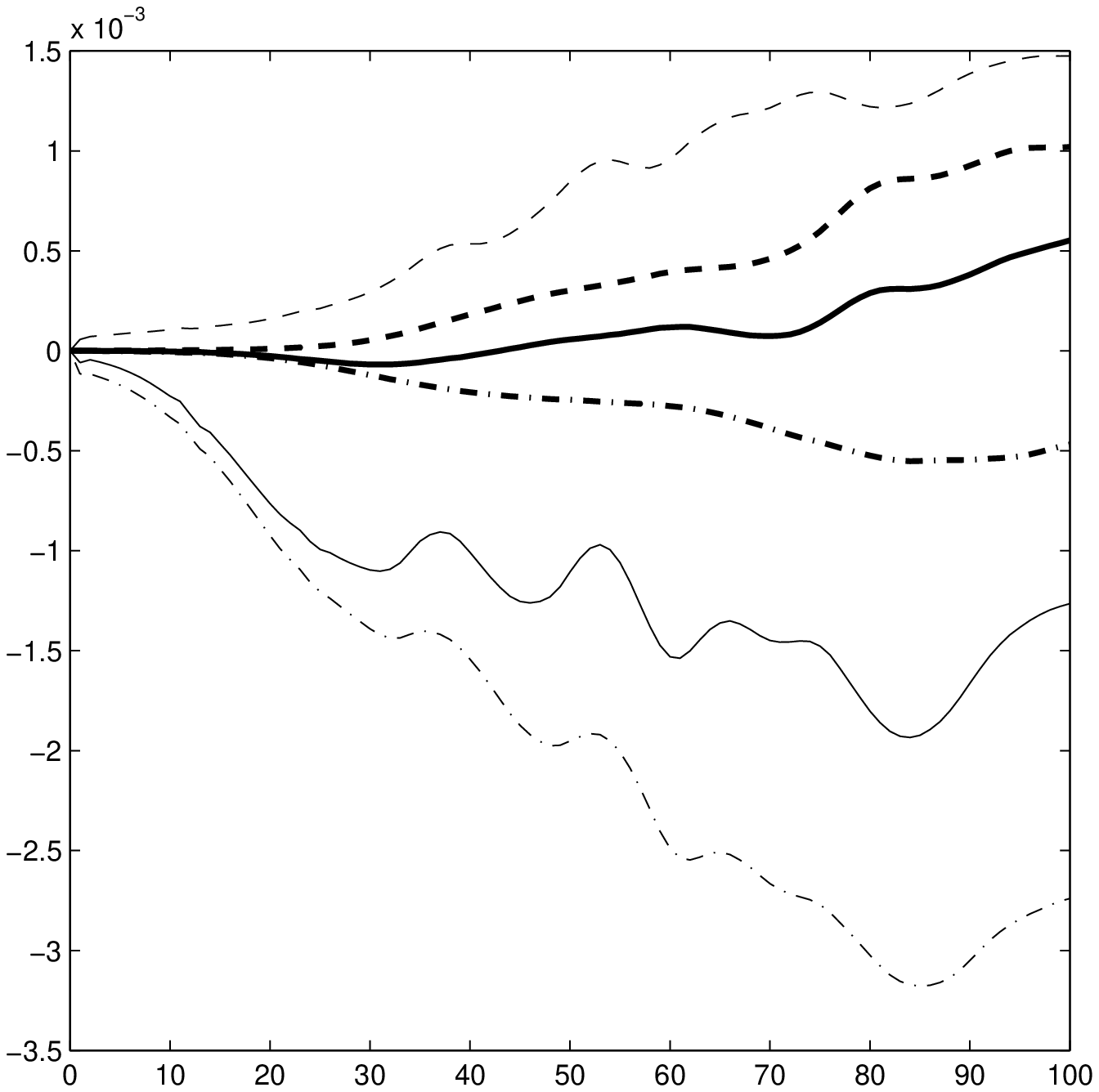,width=0.45\textwidth}(b)
}

\caption{Total ($T_t$), forward ($T_f$), backward ($T_b$) energy transfer for the LANS$-\alpha$ model comparing three different grids in (a): thin lines correspond to $32^3$:
$T_b$ dash-dotted, $T_f$ dashed, $T_t$ solid, thick lines correspond to $64^3$: $T_b$ dash-dotted, $T_f$ dashed, $T_t$ solid and markers to $96^3$: $T_b$ $\circ$, $T_f$
$\diamond$, $T_t$ $\Box$. In (b) we compare the fine-grid results for the Leray and LANS$-\alpha$ model in (b): Leray (thick lines) and LANS$-\alpha$ (thin lines): $T_t$ (solid),
$T_f$ (dashed) and $T_b$ (dash-dotted). A fixed filterwidth of $\ell/16$ was used and the Reynolds number $Re=50$.}

\label{ekin_dyn_re_50_alpha}

\end{figure}
}

The type of contribution of the LANS$-\alpha$ model to the evolution of the resolved kinetic energy is displayed in figure~\ref{ekin_dyn_re_50_alpha}. Although the convergence toward the grid-independent solution is not as dramatic as was observed in relation to the Leray model in figure~\ref{ekin_dyn_re_50} the strong backscatter effects of the LANS$-\alpha$ model are well established. We observe that the total energy transfer $T_{t}$ is negative which is an essential difference compared to the Leray model. This is best
illustrated in figure~\ref{ekin_dyn_re_50_alpha}(b) where the kinetic energy transfers of the Leray and LANS$-\alpha$ models are compared. We observe that the levels of forward and backward scatter are considerably more prominent in the LANS$-\alpha$ model. Moreover, the Leray model displays a slight negative total energy transfer only in the earlier stages of
the flow while the LANS$-\alpha$ model shows a total backscatter during the entire evolution. This is the underlying reason for the earlier observed higher level of variability in
the LANS$-\alpha$ flow.

The increased small-scale contributions in the LANS$-\alpha$ predictions require appropriate spatial resolution. The corresponding resolution requirements are discussed in some
more detail next.

\subsection{Resolution requirement for the LANS$-\alpha$ model.}
\label{alpharesolution}

{
\begin{figure}[htb]

\centerline{
\psfig{figure=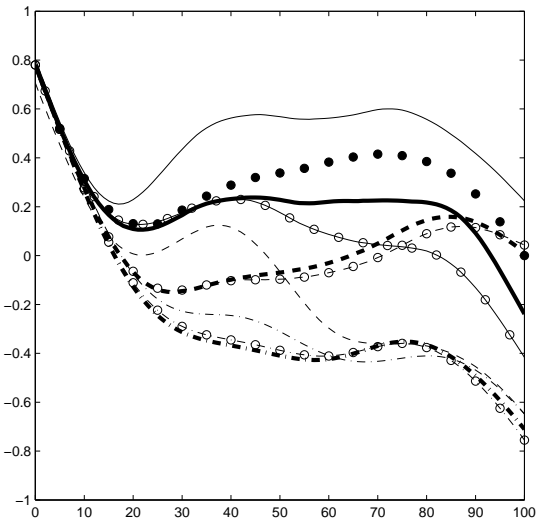,width=0.55\textwidth}
}

\caption{Convergence of momentum thickness $\delta$ at $Re=50$ for $\Delta=\ell/16$ and three LES models: LANS$-\alpha$ (solid), Leray (dashed) and
dynamic model (dash-dotted), compared with filtered DNS (solid circles). We plotted $\varepsilon(t)=\delta_{LES}(t) -\delta_{{\overline{DNS}}}(t^{*})(t/t^{*})$ for
{\mbox{$t^{*}=100$}}: $32^{3}$: no markers, $64^{3}$: open circles and $96^{3}$: thick lines.}

\label{figmomthiconv}

\end{figure}
}

To more clearly express the deviations in the momentum thickness compared to filtered DNS data, one may choose to amplify the total simulation errors that arise. We observe that the mixing layer displays a nearly linear, self-similar dependence of $\delta$ as function of time. To test this linearity and hence the convergence process more precisely, we introduce
\begin{equation}
\varepsilon(t)=\delta_{LES}(t)-\delta_{{\overline{DNS}}}(t^{*})\Big(\frac{t}{t^{*}}\Big)
\end{equation}
and use as reference time $t^{*}=100$ which corresponds to the final time used in the simulation. In figure~\ref{figmomthiconv} we consider the convergence of the predictions as a function of spatial resolution for the three subgrid models adopted in this paper. We observe that the dynamic model results are less accurate; but the solution is numerically captured to its full potential already at $32^{3}$. The Leray predictions are seen to require $64^{3}$ cells in order to attain their full potential for accurate prediction. The results of the LANS$-\alpha$ model are slightly more sensitive and approach grid-independence only at $96^{3}$. This test of linearity of $\delta$ also illustrates the consequences of under-resolution. Although the LANS$-\alpha$ predictions are the most accurate among the models at proper subgrid resolution, the effects of numerical contamination at insufficient resolution can be strong enough to lose most of this potential, e.g., on a grid with $32^{3}$ cells. This pattern of dependency of its predictions on the spatial resolution when the LANS$-\alpha$ model is under-resolved was also observed for other flow properties, such as the decay of the resolved kinetic energy.

From these numerical illustrations we may infer a rule for the spatial resolution requirements of the LANS$-\alpha$ model. As discussed in subsection~\ref{regclose} the LANS$-\alpha$
model derives its name from the length-scale parameter $\alpha$. By considering the Taylor expansion of the top-hat filter-operation and comparing this with the inverse
Helmholtz operator one arrives at the identification {\mbox{$\alpha\approx \Delta/5$}} \cite{geurtsholm2002}. Combined with the observation made above that a reliable treatment
of the LANS$-\alpha$ model requires a subgrid resolution $r \approx 6$ leads to the conclusion that LES of turbulent mixing based on the LANS$-\alpha$ model requires $\alpha \approx
h$.

\section{Concluding remarks}
\label{concl}
\subsection{Summary of results}
In this paper, we proposed to consider the mathematical approaches for regularisation of the Navier-Stokes equations as models for parameterising the effects of the unresolved length-scales in large eddy simulation of turbulent mixing flow. Two related regularisation principles were considered, namely, the Leray approach and the LANS$-\alpha$ approach. These regularisation approaches produced accurate LES predictions that do not depend on additional ad hoc implementation steps, such as those required in implementing the dynamic procedure. The LANS$-\alpha$ model is a deformation of the Leray model which recovers the Kelvin circulation theorem of the Navier-Stokes equations, but for a material loop moving with a filtered transport velocity. In a simple Fourier analysis, these regularisation models were compared to Bardina's similarity model and to the nonlinear tensor-diffusivity model. Expansions of the turbulent stress tensor provided a point of reference for these comparisons. A more meaningful assessment of the quality of these subgrid models was obtained by comparing them in numerical simulations of turbulent transport in a temporal mixing layer. At the level of instantaneous solutions, the regularisation models were shown to correspond much better with filtered DNS data than has been seen for other, more traditional subgrid models in literature. This general impression also translates into a better prediction of various flow properties, ranging from mean flow to spectral quantities.

The Leray model was shown to provide a representation of turbulent mixing that is more accurate in many respects than the predictions obtained using the dynamic
eddy-viscosity model. Moreover, on any simulation grid the computational effort required for the Leray model is considerably lower than required for the dynamic model. The
resolved kinetic energy was found to be slightly overestimated by the Leray model, but was otherwise very similar to that for the dynamic model. Important improvements were observed in the capturing of the momentum thickness and the velocity fluctuation profiles. Particularly, the intermediate and smaller resolved scales in the turbulent regime were much better represented by the Leray model, compared to the dynamic model. In addition, the intricacies of turbulent energy dynamics are contained more fully in the Leray model, because it allows both forward and backward scatter of energy. Finally, the Leray model was found to be robust at very high Reynolds number and the prediction of the dominant part of the resolved kinetic energy spectrum was found to approach a $-5/3$ behaviour with increasing Reynolds number.

Compared to the Leray and the dynamic subgrid models, the LANS$-\alpha$ model was shown to yield significant improvements, provided the spatial resolution is such that $\alpha \approx h \approx \Delta/5$. The improvements associated with the grid-independent solution displayed a rather strong deterioration in cases where the resolution was not adequate. It was observed that at subgrid resolution ratio $r=\Delta/h=2$ the accuracy of the LANS$-\alpha$ predictions decreased to being the same as that of the Leray and dynamic models. The requirement of adequate subgrid resolution posed some limitations on the practical use of the regularisation models. We concluded for the Leray model that $r \gtrsim 4$ and for the LANS$-\alpha$ model that $r \gtrsim 6$ constitute reliable values for the specification of the spatial resolution. 

\subsection{Comparison of the regularisation models and outlook}
The Leray model displays excellent robustness with increasing Reynolds number. This feature allows one to apply the Leray model accurately at reasonable computational costs and under flow-conditions that are well outside current DNS capabilities. However, the LANS$-\alpha$ model yields solutions with more realistic variability, corresponding better to the filtered DNS results than for the Leray model. Thus, a trade-off emerges between these two models. The solutions of the LANS$-\alpha$ model may more accurately represent the effects of intermittency in turbulence than the less-variable solutions of the Leray model. However, the LANS$-\alpha$ model is less robust and its application to flow at high Reynolds numbers is not as straightforward as with the Leray model. Further  investigation of this trade-off may lead to interesting developments in the comparison of the time-dependent solutions of these two models.

A convenient benefit of the regularisation approach to turbulence modelling is that it enables one to derive the implied small-scale treatment from the underlying regularisation principle. This yields a systematic closure of the equations whose analysis allows an extension in which the filter width $\Delta$ is determined dynamically by the evolving flow. The evolving filter-width may even be anisotropic \cite{Ho1999,Ho2002}. The application of this self-adaptive modelling approach in a spatially developing mixing layer and, more importantly, in near wall turbulence is a topic of current research.

\section*{Acknowledgement}

BJG gratefully acknowledges support from the Turbulence Working Group (TWG)at the Centre for Non-Linear Studies (CNLS) at Los Alamos National Laboratory which facilitated an
extended research visit in 2004 and allowed many fruitful discussions with members of TWG. Simulations were performed at SARA and made possible through grant SC-244 of the Dutch
National Computing Foundation (NCF).

\end{document}